\documentclass[usenatbib]{mn2e}

\usepackage{psfig}
\usepackage{rotating}

\title[Properties of FIRBACK-ELAIS $175\mu m$ sources in
the ELAIS N2 region]
  {Properties of FIRBACK-ELAIS $175\mu m$ sources in
the ELAIS N2 region }
\author[E. L.~Taylor et al.]
  {E. L.~Taylor$^1$, R. G.~Mann$^1$, A. N.~Efstathiou$^2$,
T. S. R.~Babbedge$^3$, \newauthor M.~Rowan-Robinson$^3$, G.~Lagache$^4$,
A.~Lawrence$^1$, S.~Mei$^5$, M.~Vaccari$^3$, \newauthor
Ph.~H\'eraudeau$^6$, S. J.~Oliver$^7$, M.~Dennefeld$^8$,
I.~Perez-Fournon$^9$, \newauthor S.~Serjeant$^{10}$,
E.~Gonz\'alez-Solares$^{11}$, J.--L.~Puget$^4$, H.~Dole$^{12}$, C.~Lari$^{13}$\\
  $^1$Institute for Astronomy, University of Edinburgh, Royal
  Observatory, Blackford Hill, Edinburgh, EH9 3HJ, UK\\
  $^2$Department of Computer Science and Engineering, Cyprus College,
6 Diogenes Str, 1516 Nicosia, Cyprus\\
  $^3$Astrophysics Group, Imperial College London, Blackett
Laboratory, Prince Consort Road, London SW7 2BZ\\
  $^4$Institut d'Astrophysique Spatiale, B\^at. 121, Universit\'e Paris
XI, F-91405 Orsay Cedex\\
  $^5$Johns Hopkins University, 3400 N. Charles Street, 21218,
Baltimore, MD, USA\\
  $^6$Kapteyn Astronomical Institute, P.O. Box 800, 9700 AV Groningen, 
The Netherlands\\
  $^7$Astronomy Centre, Department of Physics \& Astronomy, University 
of Sussex, Brighton, BN1 9QJ, UK\\
  $^8$Institut d'Astrophysique de Paris, 98bis Boulavard Arago,
F-75014 Paris\\
  $^9$Instituto de Astrofisica de Canarias, C/V\'ia L\'actea s/n. 38200 La Laguna, Tenerife, Spain\\
  $^{10}$Centre for Astrophysics \& Planetary Science, School of
Physical Sciences, University of Kent, Canterbury, Kent CT2 7NR\\
  $^{11}$Institute of Astronomy, University of Cambridge, Madingley
Road, Cambridge CB3 0HA\\
  $^{12}$Steward Observatory, University of Arizona, 933 N Cherry Ave
Tuscon AZ 85721, USA\\
  $^{13}$Instituto do Radioastronomia, Via P. Gobetti 101, 40129,
Bologna, Italy}

\date{Received ??}

\begin{document}

\maketitle

\begin{abstract}
We report on a search for the optical counterparts of 175 ${\rm \mu m}$ -- selected sources
from the Far--Infrared Background (FIRBACK) survey in the European
Large Area ISO Survey (ELAIS) N2 field. Applying a likelihood
ratio technique to optical catalogues from the Isaac Newton Telescope
-- Wide Field Survey (INT--WFS), we found optical identifications for
33 out of 55 FIRBACK sources in this field. These were
then reassessed in the light of associations with the ELAIS final
catalogue for the the N2 field, to yield a final set of 31
associations. We have investigated the nature of
this population through a comparison of their observed spectral energy
distributions (SEDs) with predictions from radiative transfer models which
simulate the emission from both cirrus and starburst components. We
find the far--infrared sources to be 80 per cent star bursting galaxies with
their starburst component at a high optical depth. The resulting
SEDs were used to estimate far--infrared
luminosities, star formation rates, dust temperatures and dust masses.
The N2 FIRBACK population is found to consist of four suspected
ultra--luminous infrared galaxies (ULIRGs)
with ${\rm L}_{{\rm FIR}}\sim 10^{12}$ ${\rm L}_{\odot}$ and
${\rm SFR}_{{\rm FIR}}>100$ ${\rm M}_{\odot}\textrm{yr}^{-1}$, a number of luminous
infrared galaxies (LIRGs) with moderate
star formation rates and ${\rm L}_{{\rm FIR}}\sim 10^{11}$ ${\rm L}_{\odot}$ and a
population of low redshift quiescently star forming galaxies. We also discuss the
implications of these results for current evolutionary models.
\end{abstract}


\section{Introduction}
Evidence available in the 1960s and 1970s from regions of active star formation in nearby galaxies, and
our own Milky Way, suggested that star forming galaxies would be strong
sources of thermal infrared radiation, and implied that regions of
dust and molecular clouds should be found within them
\citep{Sunyaev78}. It was then predicted that this dust reprocessed
redshifted starlight from distant galaxies, would form a far--infrared
background (FIRB) signal \citep{Partridge67}.

With the advent of the Infrared Astronomical Satellite (IRAS) in the 1980s hundreds of previously undetected 
galaxies were discovered, which emit up to 95 per cent of their total
luminosity in the infrared \citep{Soifer87}. This indicated
significant dust reprocessing of their starlight and implied that
star formation rates previously calculated from rest frame optical/uv
luminosities were only a lower limit. Follow up redshift surveys, such 
as QDOT \citep{Lawrence99}, showed that these galaxies were far more 
numerous in the past (e.g. \citealt{Oliver92}), requiring stronger cosmological evolution than
had previously been predicted by conventional models for the passive
evolution of galaxies and providing support was building for models in which
starbursts are important evolutionary mechanisms \citep{Lonsdale90}.

The IRAS results allowed far tighter limits to be put on the level of the
predicted far--infrared background. \citet{Hacking91} predicted that
the background should be detectable by the Cosmic Background Explorer
satellite (COBE) at 100 ${\rm \mu m}$ and noted
that discovering its exact level would allow further 
refinement of evolutionary models. No-evolution models placed a lower 
limit on the predicted background level, while detection of a strong
background would indicate the existence of a
population of objects with significant cosmological evolution. The
FIRB was finally discovered by \citet{Puget}
 in the whole sky
survey taken by FIRAS on COBE
just as \citet{Hacking91} had predicted. 

\citet{Puget} found a signal which could be explained by the thermal
emission from particles with temperatures of 20 and 12 K with
emissivities of $3\times 10^{-6}$ and 5 K with an emissivity of
$3\times 10^{-5}$. This detection has fueled a new generation of galaxy 
evolution models but it is clear that to further differentiate between 
them we must understand the nature of the objects which contribute to
the far--infrared background.

\subsection{The Far--Infrared Background} \label{sec:FIRB}
The FIRBACK survey was carried out jointly
with the ELAIS team at
175 ${\rm \mu m}$ using ISOPHOT \citep{Lemke96} on the Infrared Space Observatory
(ISO) \citep{Kessler}. This survey aimed to resolve the FIRB
into discrete sources and also allow studies of FIRB
fluctuations. FIRBACK covered 3 main fields [South Marano, N1 and N2] covering a total area of 4
deg$^2$. The final source catalogue contains 106 sources with
fluxes between 180 mJy and 2.4 Jy and a supplementary catalogue
containing 90 sources with fluxes 135-180 mJy
\citep{Dole99,Dole01}. The northern fields lie within the previously
defined ELAIS N1 and N2 survey regions whose selection was based
primarily on their low Galactic cirrus emission (as quantified by IRAS
100 ${\rm \mu m}$ emission) and high visibility, given ISO's orbit. This choice of fields
allows multiwavelength follow--up thus providing a global view of
galaxy evolution. 

There have been several suggestions as to the nature of these 175
${\rm \mu m}$ ISO sources. Two sources were examined in detail in the optical and
near-IR by \citet{Chapman} 
who found two nearby ($z<1$) galaxies
with characteristic temperatures of $\sim$30 K and $\sim$50 K and
bolometric luminosities (40--200 ${\rm \mu m}$) of
$10^{11}-10^{12}$ ${\rm L}_{\odot}$. These galaxies were classified as
ULIRGs with
morphologies suggesting the early stages of mergers adding weight to arguments
for interactions sparking periods of increased star formation. A sample of
ISO 175 ${\rm \mu m}$ sources were also examined at sub--mm and near-IR wavelengths by
 \citet{Sajina} 
who found a bimodal galaxy population: one
population of normal star forming galaxies at $z\sim 0$ and a second
more luminous population at $z\sim 0.4-0.9$. There has been
spectroscopic follow--up of the brightest 175 ${\rm \mu m}$ sources in the FIRBACK
South Marano field \citep{Patris03} which were found to be nearby ($z<0.3$), heavily
extincted ($A_v\sim 3$ with extreme cases up to $A_v=7$),
star forming galaxies with moderate star formation rates (a few
10 ${\rm M}_{\odot}{\rm yr}^{-1}$). These were classified at LIRGs,
with ${\rm L}_{{\rm IR}}\simeq 10^{11}$ ${\rm L}_{\odot}$. 

We have gone one step further by seeking associations for all 55 N2
$175 {\rm \mu m}$ ISO sources in the FIRBACK survey
with optical sources from the INT--WFS catalogues in four bands
(\textit{g',r',i',Z}), the other ELAIS survey wavelengths ({\it J, H, K}, 15, 6.7,
90 ${\rm \mu m}$, 20 cm) and IRAS wavelengths (60, 100 ${\rm \mu m}$). (Also 
see \citealt{Dennefeld} for investigations into sources in the N1 field). In
doing so we are able to gain a multiwavelength view of the nature of
these interesting objects. We have compared their SEDs with those predicted by the radiative transfer
models of \citet{ERR03} 
to provide further insight into their properties.

The layout of this paper is as follows: Section \ref{sec:data}
describes the multiwavelength data, Section \ref{sec:IDs} discusses the
methods of associating the multiwavelength data with the 175 ${\rm \mu m}$
ISO sources, provides a summary of the results of this process and
briefly describes the optical properties
of the sources. The radiative transfer models are described in Section 
\ref{sec:models} and 
Secton \ref{sec:SEDs} describes the comparison of the sources' SEDs with
predictions from the models. The resulting
star formation rates, far--infrared luminosities and dust temperature
estimations are given in Sections \ref{sec:sfrs} and \ref{sec:dust}.
Section \ref{sec:evolution} provides a summary
of current evolutionary models, while conclusions and a discussion are
provided in Section \ref{sec:conc}. An Appendix provides details of
individual associations.


\begin{figure}
\centerline{\psfig{figure=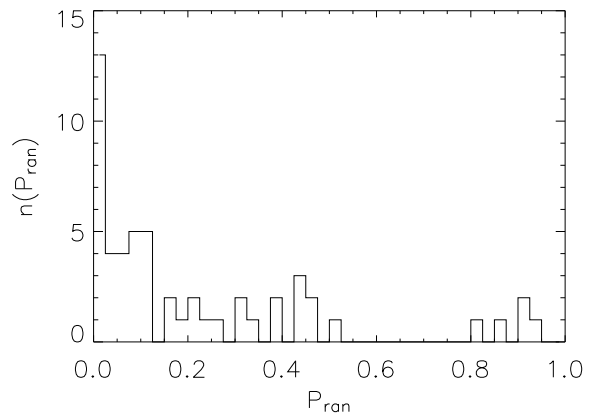,width=0.5\textwidth}}
\caption{\label{fig:like_pran} Distribution of
${\rm P_{ran}}$ values for chosen source counterparts.}
\end{figure}


\begin{figure}
\centerline{\psfig{figure=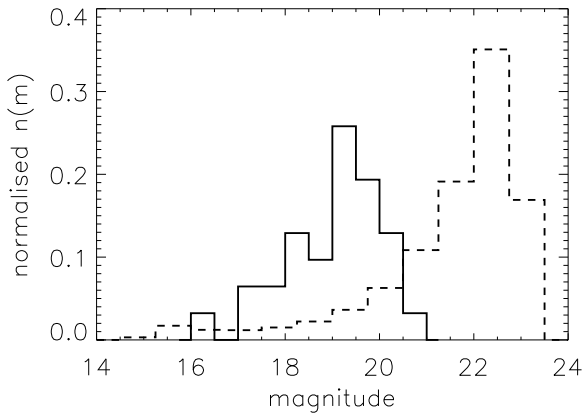,width=0.5\textwidth}}
\caption{\label{fig:mag_hist} Normalised magnitude distribution of the
N2 175 ${\rm \mu m}$ ISO source
optical counterparts ({\it r'}--band) (solid line), dashed line gives the
{\it r'}--band magnitude
distribution of the INT-WFS catalogue for comparison.}
\end{figure}


\begin{figure}
\centerline{\psfig{figure=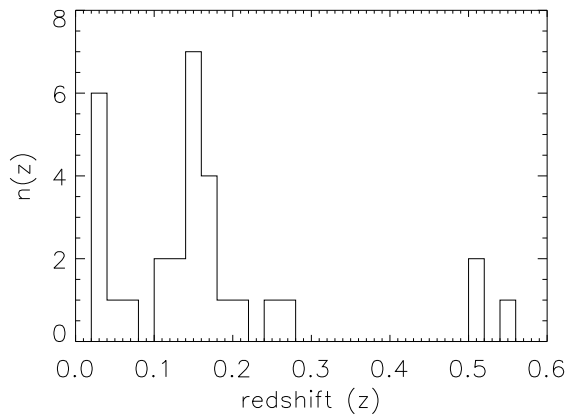,width=0.5\textwidth}}
\caption{\label{fig:z_dist} Redshift distribution for the 30 175 ${\rm
\mu m}$ sources with spectroscopic or photometric redshifts as listed in
Table 1.}
\end{figure}

\section {Complementary Data} \label{sec:data}

Optical associations were sought in \textit{r'} data from the INT--WFS 
\citep{McMahon01} which used the Wide Field
Camera (WFC) on
the 2.5 m Isaac Newton Telescope at the Observatorio del Roque de Los 
Muchachos (La Palma, Spain). The WFC is formed by four 4k x 2k
CCDs. The arrays have 13.5 ${\rm \mu m}$
pixels corresponding to 0.33 arcsec/pixel at the telescope prime focus
and each one covers an area on the sky of 22.8 arcmin x 11.4 arcmin.
The total sky coverage per exposure for the array is therefore
0.29 deg$^2$. Gaps between detectors are typically 20 arcsec and optical
observaions are carried out allowing for a 10 per cent overlap between adjacent
pointings for photometric purposes. 
The  ELAIS-N2 optical field is approximately 9 deg$^2$. The survey has
a  typical
completeness limit of 23 mag (Vega) in wavebands \textit{U, g', r', i'} and \textit{Z}. However, there
are no {\it U}--band data for our sources the N2 field. All optical magnitudes were
calculated directly from the INT--WFS data.

Associations were then sought with sources in the multiwavelength
ELAIS band-merged catalogue at
6.7, 15 and 90 ${\rm \mu m}$ with associated data at
\textit{U,g',r',i',Z,J,H,K} and 20 cm \citep{ELAIS} which by then had
become available.  This was a survey
carried out using ISO 
covering a total of 12 deg$^2$ over five main fields, three in the
north (N1, N2, N3) and two in the south (S1,S2). These areas were also
surveyed at 20 cm with the Very Large Array (VLA) and AT. The separate wavebands making
up the ELAIS survey each comprise an independent survey, the
band--merged catalogue simply brings together the results of each of
their final anayses, as described in Section
\ref{sec:IDs}.

Where detections
were not found at 15 and 90 ${\rm \mu m}$ upper limits were extracted
directly from the survey maps (\citealt{15um,90um}) to a $3\sigma$ level using aperture
photometry with an aperture size of 6 arcsec and 90 arcsec respectively. An
aperture correction was applied to the 15 ${\rm \mu m}$ upper limits since
40 per cent of the point spread function (PSF) lies outside the aperture.

IRAS 100 and 60 ${\rm \mu m}$ fluxes or upper limits
($3\sigma$) were obtained at the galaxy optical positions 
using the IRAS Scan Processing and Inegration (SCANPI) internet facility\footnote{http://irsa.ipac.caltech.edu/applications/IRAS/Scanpi/} to
provide extra constraints on the source SEDs.

Associations were also sought in the Submillimetre Common User
Bolometer Array (SCUBA) 850 ${\rm \mu m}$  \citep{Scott02} and Max--Planck Millimeter
Bolometer array (MAMBO) 1200 ${\rm \mu m}$ \citep{Greve04} catalogues
within 10 and 3 arcsec of the optical position of the 175 ${\rm \mu m}$ counterpart, respectively.
However, none were found for any of our 175 ${\rm \mu m}$ sources.


\begin{figure*}
\centerline{\psfig{figure=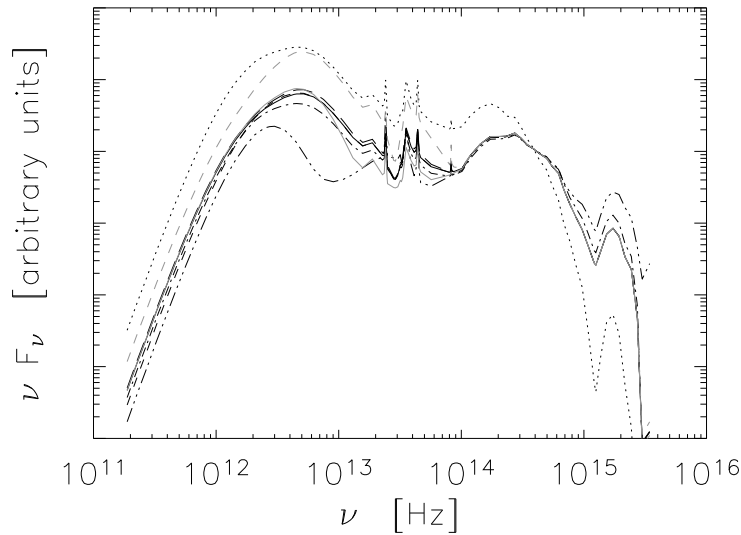,width=0.65\textwidth}}
\caption{\label{fig:mod_examples} Examples of the affects of the
radiative transfer model
parameters on the resulting SEDs, see Section \ref{sec:models}. Solid
black line: $A_V=1$, $f_{SB}=1$, $\tau_v=100$ (comparison model). Black dotted line:
$A_V=3$, $f_{SB}=1$, $\tau_v=100$. Dash-dot-dot-dot line: $t_m=0$.
Grey dashed line: $A_V=1$, $f_{SB}=5$, $\tau_v=100$. Dashed line:
$\psi=10$. Dash-dot line: $\tau=8\times 10^{10}$. Long dashed line:
$\tau_{SB}=80$ Myrs. Solid grey line: $\tau_v=200$.}
\end{figure*}

\section{Identifications: method and results} \label{sec:IDs}

The positional accuracy of the ISO 175 ${\rm \mu m}$ sources was
determined using simulated maps by \citet{Dole01}. Since
93 per cent of artificial sources brighter than 180 mJy in the simulated maps were recovered
within a radius of 50 arcsec, associations were sought for the N2
sources within a 90 arcsec radius. This ensured the capture within the
source error circle of the true source position for even faint source
IDs. Associations were made using a likelihood ratio technique
described in full in \citet{Mann}. 
In brief, the likelihood ratio
is the ratio of the probability of finding the true counterpart to a
particular source at the source position with that magnitude to the
probability of finding an object of the same magnitude there by
chance. In terms of flux this takes the form:
\begin{equation}
{\rm LR=\frac{q(f)e(x,y)}{n(f)}}
\end{equation}
\noindent where q(f) is the flux probability distribution function for
the source counterparts, e(x,y) is the probability distribution of positional offsets
between the source and object and n(f) is the surface density of
objects per unit flux interval. e(x,y) was taken to be a gaussian
distribution with $\sigma =20$ arcsec. This choice prevents too many
objects being assigned moderate likelihood ratios, without excluding
objects which should be considered or including objects which should
not. The function q(f) is unknown for a new
source population, as is the case here, and, as shown in
 \citet{Mann2}, an empirical estimate is very noisy for a small
number of sources, q(f) was taken to be a constant. The probability,
${\rm P_{ran}}$, that a fictitious, randomly-placed source would have a likeliest
assocation with an optical object in the catalogue producing a
likelihood ratio at least as high as the source-object pair in question
was then calculated. Thus good associations have low values of ${\rm P_{ran}}$,
corresponding to a high probability that the association is
correct. The object with the lowest ${\rm P_{ran}}$ was chosen as the
source optical ID, with a cut off for non-association of
${\rm P_{ran}}=0.15$. This is an arbitrary value chosen by inspection of the
${\rm P_{ran}}$ distribution, see Figure \ref{fig:like_pran}. It is expected 
\citep{Mann} that true IDs will cluster at low ${\rm P_{ran}}$ values with a tail of
objects reaching to high ${\rm P_{ran}}$ values due to
sources without good associations in the optical catalogue, as seen in
the plot.

Counterparts were also sought in the band-merged ELAIS multiwavelength
catalogue \citep{ELAIS} using a search radius of 60 arcsec consistent, with the band
merging procedure used in the collation of this catalogue.  The
individual wavelength ELAIS catalogues were merged, taking into account the
different positional accuracies of each catalogue,
sequentially as follows. The 15 ${\rm \mu m}$ and 20 cm catalogues \citep{15um,Ciliegi99} were
first identified with optical objects from the INT--WFS catalogue \citep{Gonzalez03} and
then merged according to their optical positions using a search radius
of 2 arcsec. These were then matched with the 6.7 ${\rm \mu m}$ catalogue using a
search radius of 5 arcsec and those 6.7 ${\rm \mu m}$ sources not merged were
associated with optical counterparts. These were then merged with the
90 ${\rm \mu m}$ and 175 ${\rm \mu m}$ sources \citep{90um,Dole01}
using search radii of 30 arcsec and
60 arcsec respectively. Where these were matched with more than one
catalogue entry the less probable assocations were flagged in the
catalogue. After this those sources that were still unmatched were
associated with optical counterparts. These optical counterparts were
sought by \citet{ELAIS} independently from our work and the resulting optical IDs were then compared.

In summary, we sought associations on the basis of applying the
likelihood ratio method to the optical data alone, then, with the
availablility of the final band-merged ELAIS catalogue, we reassessed
them. Our final optical and ELAIS wavelength ID list was drawn up
using the following criteria:
\begin{enumerate}
\item If our optical ID, selected using the likelihood ratio method,
agreed with the ELAIS optical ID to
which the 175 ${\rm \mu m}$ flux had been assigned then both the ELAIS
optical ID and band-merged data were accepted.
\item If the ELAIS optical ID had a ${\rm P_{ran}}<0.15$ (cut off value see
earlier) for association with the 
175 ${\rm \mu m}$ ISO source then the ELAIS
optical ID and band-merged data were accepted.
\item If our optical ID and ELAIS optical ID were in fact the
same galaxy but had different entries in the INT--WFS catalogue due to the 
way the image analyser splits up bright galaxies into multiple sources, 
then the band-merged ELAIS data and optical ID were accepted. 
\item If our optical ID had ${\rm P_{ran}}<0.15$ but no
optical ELAIS catalogue match then the source has optical and 175 ${\rm \mu m}$ data
only (it may also have IRAS detections/upper limits and 60 ${\rm \mu m}$ and
90 ${\rm \mu m}$ upper limits)
\item If our optical ID had ${\rm P_{ran}}>0.15$ the source
has no association
\end{enumerate}

For sources in the main N2 FIRBACK catalogue ($S>180$ mJy), which
contains 27 detections, we made 22 confident optical associations
with galaxies, 7 of which agree exactly with the ELAIS optical
ID (category i.) and 6 for
which we accepted the ELAIS optical ID (categories ii. or iii.). Nine
confident optical associations were made for sources in the supplementary catalogue
($S<180$ mJy, 28 sources), 2 of which agreed with the ELAIS
optical ID 
(category i.) and 2 of
which where the ELAIS optical ID was accepted (categories ii. or iii.). There also were two sources in the
supplementary catalogue with stellar IDs, one of which agreed with the 
ELAIS optical ID (category i.) and one of which was an
accepted ELAIS optical ID
(category ii. or iii.). The data for the identified sources are given in
Table 1. and notes
on individual sources are given in the Appendix, including ${\rm P_{ran}}$
values of any other plausible identifications.

The identified sources were morphologically classified
by visual inspection. We used \textit{i'}-band images from the INT--WFS, and followed the same
classification method as \citet{Postman05}. The galaxies were
classified using the classical Hubble sequence: E, E/S0, S0, S0/a, Sa,
Sa/b, Sb, Sb/c, Sc, Irr. For the Advanced
Camera for Surveys (ACS) on HST, for galaxies with redshift of around unity, visual classifications of this kind
have a typical random error of 25 per cent, this reduces to 6 per cent
when galaxies are classified in two broad cathegories: 1. early-type,
2. spirals and irregulars. These errors were estimated from the
average scatter found in classifications by four ACS
team members. The majority of our sources are spiral
galaxies, we also have a number of S0 galaxies, 2 sets of interacting
galaxies, 2 pairs and 2 irregulars, the morphology of 6 sources was
unable to be deduced from the images.

\begin{sidewaystable*}
\scriptsize
\label{tab:data}
\begin{tabular}{@{}lccc@{  }c@{  }c@{  }c@{  }c@{  }c@{  }c@{  }c@{  }c@{  }c@{  }c@{  }l@{}}
\multicolumn{14}{l}{{\bf Table 1.} Multiwavelength data for associated sources, fluxes in mJy.}\\ 
\multicolumn{14}{l}{nb. Sources listed with ${\rm P_{ran}}$ greater than the
threshold 0.15 due to falling into category (iii) (see Section \ref{sec:IDs})}\\
\multicolumn{14}{l}{FN2 004 has ${\rm P_{ran}}=1.0$ since the
associated galaxy is mistakenly classified as a star in the INT WFS catalogue.}\\
\hline
FIRBACK & FIRBACK position & optical position & 175 ${\rm \mu m}$ &
${\rm P_{ran}}$ & {\it r'}--mag & 15 ${\rm \mu m}$ &
6.7 ${\rm \mu m}$ & 60 ${\rm \mu m}$ & 90 ${\rm \mu m}$ & 100 ${\rm
\mu m}$ & 20 cm & $z_{phot}$ &
$z_{spec}$ & morphological classification\\
\hline
 FN2 000 & 16 37 33 +40 52 26 &16 37 34.53  +40 52 11.2  & 2377$\pm$213 &0.026
& 18.46 & 53.9$\pm$0.3 &  18.3$\pm$0.1  &  - &  1461$\pm$36 & - & 8.74$\pm$0.02 & - & 0.03 &
irregular or edge--on \\
 &&&&&&&&&&&&&&spiral with tidal tail\\
 FN2 001 & 16 35 08 +40 59 20 &16 35 07.87  +40 59 28.9  & 1251$\pm$139. &
0.002 & 18.28 & 26.1$\pm$0.4 &  4.11$\pm$0.12 &  0.15$\pm$0.03 & 337$\pm$15 &   -  &  0.76$\pm$0.04 & 0.10 &
0.03 & edge--on\\
 &&&&&&&&&&&&&&Sb/c\\
 FN2 002 & 16 36 10 +41 05 16 &16 36 08.15  +41 05 07.7  &  803$\pm$102 &
0.086 & 19.33 &  8.9$\pm$0.1 &  2.3$\pm$0.1 &  0.41$\pm$0.02 & 614$\pm$37 &   -  &   -  & 0.15 &
0.17 & irregular with\\
 &&&&&&&&&&&&&&tidal tail\\
 FN2 003 & 16 35 25 +40 55 51 &16 35 25.22  +40 55 42.1  &  682$\pm$92 &
0.001 & 17.45 & 13.9$\pm$0.1 &  5.5$\pm$0.1 &  0.19$\pm$0.03 & 416$\pm$19 &   -  &  1.70$\pm$0.02 & 0.17 &
0.03 & Sa\\
 FN2 004 & 16 34 01 +41 20 49 &16 34 01.82  +41 20 52.5  &  666$\pm$91 &
1.000 & 16.41 & 20.4$\pm$0.1 &  9.6$\pm$0.1 &  0.40$\pm$0.02 & 403$\pm$26 &   -  &  1.8$\pm$0.02 & 0.05 &
0.03 & SBa\\
 FN2 005 & 16 32 43 +41 08 38 &16 32 42.39  +41 08 46.1  &  522$\pm$78 &
0.011 & 18.67 &  6.1$\pm$0.2 &  2.1$\pm$0.1 &  0.16$\pm$0.03 & 399$\pm$27 &   -  &  0.67$\pm$0.04 & 0.02 &
0.26 &Sa\\
 FN2 007 & 16 35 45 +40 39 14 &16 35 46.91  +40 39 03.4  &  316$\pm$60 &
0.022 & 17.95 &  6.1$\pm$0.2 &  2.4$\pm$0.1 &  0.09$\pm$0.03 &  84$\pm$25 &   -  &  0.92$\pm$0.02 & 0.02 &
0.12 & compact source \\
 &&&&&&&&&&&&&&with 2 nuclei\\
 FN2 008 & 16 35 47 +41 28 58 &16 35 48.04  +41 28 30.3  &  293$\pm$58 &
0.084 & 18.32 &  4.5$\pm$0.1 &  1.3$\pm$0.1 &  0.12$\pm$0.03 &  74$\pm$18 &   -  &  0.38$\pm$0.01 & 0.10 &
0.14 & Sb\\
 FN2 010 & 16 35 38 +41 16 58 &16 35 36.16  +41 17 27.3  &  285$\pm$57 &
0.451 & 19.47 &  2.8$\pm$0.6 &   -  &  0.06$\pm$0.03 &  74.00$\pm$17 &   -  &  0.80$\pm$0.01 & 0.02 &
0.17 & pair: irregular with \\
 &&&&&&&&&&&&&&tidal tail and Sa\\
 FN2 011 & 16 38 07 +40 58 12 &16 38 08.78  +40 58 07.4  &  260$\pm$55 &
0.118 & 19.78 & $<$1.97 &   -  & $<$0.12 &  $<$0.44 & $<$0.26 &   -  & 0.51 &
- & Sa?\\
 FN2 012 & 16 34 13 +40 56 45 &16 34 11.98  +40 56 52.8  &  249$\pm$54 &
0.088 & 19.99 &  2.7$\pm$0.1 &  1.0$\pm$0.1 & $<$0.08 &  83$\pm$10 &   -  &  0.51$\pm$0.01 & 0.07 &
0.14 & Sa?\\
 FN2 015 & 16 36 07 +40 55 37 &16 36 07.71  +40 55 47.1  &  223$\pm$51 &
0.015 & 18.64 &  5.5$\pm$0.1 &  1.2$\pm$0.1 & $<$0.06 & 109$\pm$11 &   -  &  0.81$\pm$0.02 & 0.26 &
0.17 & interacting galaxies: \\
 &&&&&&&&&&&&&& Sa/b and Sa\\
 FN2 016 & 16 34 26 +40 54 07 &16 34 23.90  +40 54 10.0  &  218$\pm$51 &
0.191 & 19.88 &  3.1$\pm$0.1 &  1.0$\pm$0.1 &  0.06$\pm$0.03 &  $<$0.25 &  0.20$\pm$0.08 &  0.55$\pm$0.01 & 0.10
& 0.13 & edge--on: \\
 &&&&&&&&&&&&&&type not clear\\
 FN2 017 & 16 34 44 +41 08 42  &16 34 44.90  +41 08 20.6  &  213$\pm$50 &
0.030 & 18.46 &  2.2$\pm$0.7 &   -  & $<$0.10 &  $<$0.25 & $<$0.18 &   -  & 0.18 &
- &interacting galaxies:\\
 &&&&&&&&&&&&&&Sa with edge--on\\
 FN2 018 & 16 33 38 +41 01 15 &16 33 37.23  +41 01 09.1  &  212$\pm$50 &
0.037 & 19.51 & $<$2.98 &   -  &  0.08$\pm$0.03 &  $<$0.28 & $<$0.19 &   -  & 0.15 &
- & S0/a\\
 FN2 019 & 16 37 17 +40 48 36 &16 37 16.80  +40 48 25.6  &  205$\pm$49 &
0.001 & 18.05 & $<$3.99 &   -  &  0.06$\pm$0.04 &  $<$0.26 & $<$0.18 &   -  & 0.10 &
0.03 & Sb\\
 FN2 020 & 16 32 41 +41 06 10 &16 32 40.50  +41 06 15.4  &  201$\pm$49 &
0.086 & 20.21 & $<$6.58 &   -  & $<$0.07 &  $<$0.25 & $<$0.11 &   -  & 0.15 &
- & -\\
 FN2 021 & 16 37 58 +40 51 21 &16 37 59.39  +40 51 15.8  &  196$\pm$49 &
0.001 & 17.57 & $<$2.18 &   -  & $<$0.07 &  $<$0.19 & $<$0.17 &   -  & 0.10 &
- & -\\
 FN2 022 & 16 37 08 +41 28 26 &16 37 08.21  +41 28 56.1  &  190$\pm$48 &
0.140 & 18.80 &  2.5$\pm$0.1 &  1.7$\pm$0.2 & $<$0.10 &  $<$0.28 & $<$0.23 &  0.56$\pm$0.02 & 0.05
& 0.17 & Sa\\
 FN2 023 & 16 33 51 +40 49 44 &16 33 51.65  +40 49 46.3  &  188$\pm$48 &
0.018 & 19.04 &  2.5$\pm$0.7 &   -  & $<$0.11 &  $<$0.28 & $<$0.32 &   -  & 0.02 &
- & Sb/c\\
 FN2 025 & 16 36 31 +40 47 38 &16 36 31.23  +40 47 24.7  &  184$\pm$48 &
0.119 & 20.24 &  2.1$\pm$0.7 &   -  &  0.11$\pm$0.03 &  $<$0.33 &  0.11$\pm$0.07 &   -  & 0.20 &
- & Sa\\
CFN2 029 & 16 34 20 +41 06 54 &16 34 19.46  +41 06 37.8  &  178$\pm$47 &
0.070 & 19.62 & $<$4.36 &   -  &  0.06$\pm$0.02 &  $<$0.21 & $<$0.24 &   -  & 0.15 &
- & -\\
CFN2 030 & 16 35 23 +40 38 42 &16 35 22.81  +40 38 37.1  &  178$\pm$47 &
0.086 & 20.26 & $<$3.66 &   -  &  0.06$\pm$0.03 &  $<$0.24 &  0.17$\pm$0.05 &   -  & 0.15 &
- & S0/a\\
CFN2 034 & 16 34 12 +40 46 26 &16 34 12.52  +40 46 34.4  &  166$\pm$46 &
0.030 & 19.42 & $<$2.61 &   -  & $<$0.08 &  $<$0.11 & $<$0.10 &   -  & 0.55 &
- & S0/a\\
CFN2 036 & 16 37 01 +40 43 08 &16 36 59.92  +40 42 46.6  &  165$\pm$46 &
0.119 & 19.40 & $<$5.08 &   -  &  0.04$\pm$0.03 &  $<$0.25 & $<$0.28 &   -  & 0.10 &
- & Sa\\
CFN2 038 & 16 34 32 +41 22 37 &16 34 31.57  +41 22 45.7  &  161$\pm$45 &
0.023 & 19.26 &  1.8$\pm$0.1 &   -  & $<$0.08 &  $<$0.17 &  0.12$\pm$0.05 & $<$0.15 & 0.02 &
0.14 & Sb/c\\
CFN2 039 & 16 36 13 +40 42 25 &16 36 13.65  +40 42 30.0  &  160$\pm$45 &
0.001 & 17.15 &  5.9$\pm$0.2 &  2.5$\pm$0.1 &  0.11$\pm$0.00004 &  90$\pm$21 &   -  &  0.88$\pm$0.01 & 0.02 &
0.07 & Sb\\
CFN2 044 & 16 37 26 +40 45 39 &16 37 25.93  +40 45 37.1  &  150$\pm$44 &
0.053 & 19.96 & $<$2.87 &   -  &  0.02$\pm$0.05 &  $<$0.30 & $<$0.14 &   -  & 0.05 &
- & S0/a\\
CFN2 047 & 16 34 51 +41 20 27 &16 34 49.54  +41 20 49.2  &  147$\pm$44 &
0.125 & 19.16 &  2.4$\pm$0.1 &  1.3$\pm$0.1 & $<$0.05 &  $<$0.20 & $<$0.21 &  0.15$\pm$0.04 & 0.07
& 0.25 & S0/a\\
CFN2 049 & 16 37 42 +41 19 11 &16 37 41.44  +41 19 14.8  &  143$\pm$44 &
0.210 & 20.87 &  1.9$\pm$0.2 &   -  &  0.09$\pm$0.03 &  74$\pm$24 &   -  & $<$0.15 & 0.51 &
- & Sa?\\
\hline
\end{tabular}
\end{sidewaystable*}

Figure \ref{fig:mag_hist} shows the optical (\textit{r'}--band) magnitude distribution of
the N2 175 ${\rm \mu m}$ ISO source optical counterparts compared to that of the
overall INT-WFS catalogue. The source IDs are, in general, bright
objects with apparent magnitudes ranging from 21 to as bright as 16.

Figure \ref{fig:z_dist} shows the redshift distribution of the sources:
they lie predominantly at $z<0.3$ with only two sources having
$z>0.5$. Redshifts are discussed in more detail in Section \ref{sec:SEDs}.


\begin{figure}
\centerline{\psfig{figure=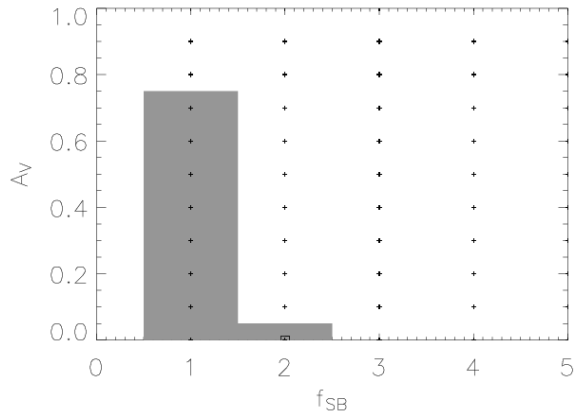,width=0.5\textwidth}} 
\caption{\label{fig:fsb_av}$1\sigma$ starburst model parameter
confidence region for
$f_{SB}$ and $A_V$ for FN2 008. Possible models parameter combinations are shown
with small crosses (note $f_{SB}=1-100$ and $A_V=0-3$ were allowed), the best-fitting model 
is marked with a square and the $1\sigma$ confidence region is
shaded. {\bf A high resolution version of this figure is available at
http://www.roe.ac.uk/$\sim$elt/N2paper\_images.html .}}
\end{figure}


\begin{figure}
\centerline{\psfig{figure=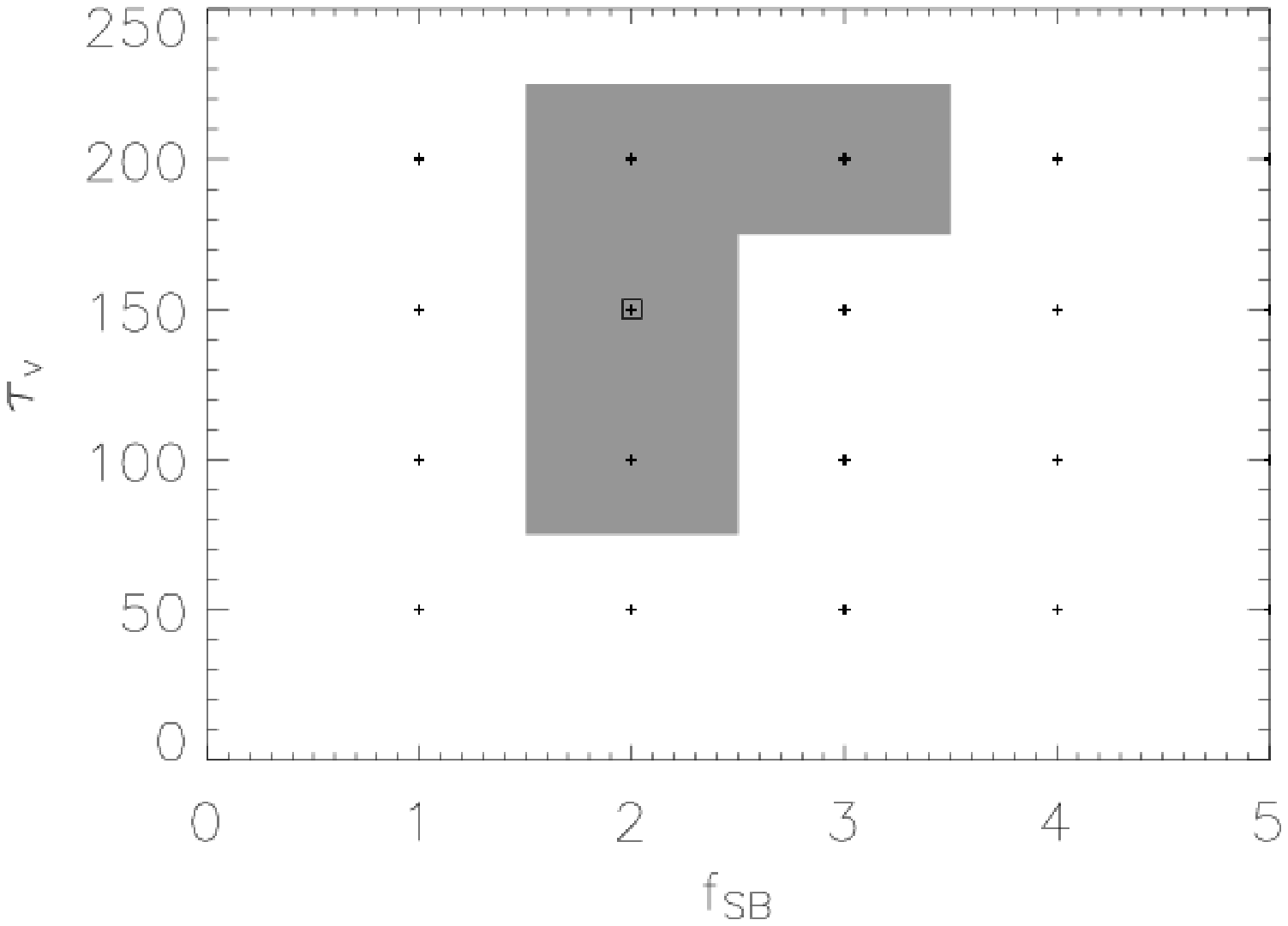,width=0.5\textwidth}}
\caption{\label{fig:fsb_tauv}$1\sigma$ starburst model parameter
confidence region for
$f_{SB}$ and $\tau_v$ for FN2 008. Possible models parameter combinations are shown
with small crosses (note $f_{SB}=1-100$ and $\tau_v=50-200$ were allowed), the best-fitting model 
is marked with a square and the $1\sigma$ confidence region is
shaded. {\bf A high resolution version of this figure is available at
http://www.roe.ac.uk/$\sim$elt/N2paper\_images.html .}}
\end{figure}


\begin{figure}
\centerline{\psfig{figure=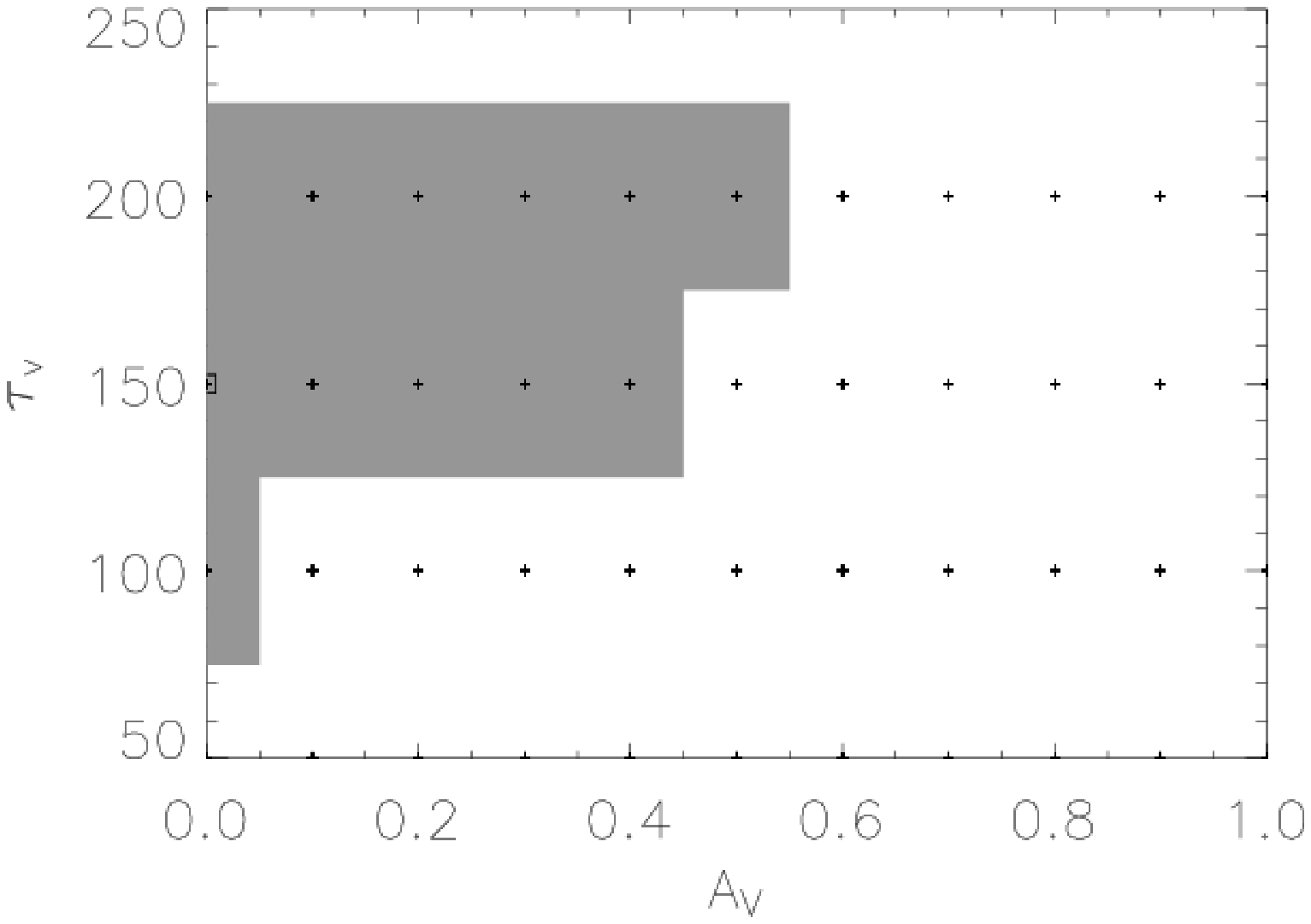,width=0.5\textwidth}} 
\caption{\label{fig:av_tauv}$1\sigma$ starburst model parameter
confidence region for
$A_{V}$ and $\tau_v$ for FN2 008. Possible models parameter combinations are shown
with small crosses (note $A_{V}=0-3$ and $\tau_v=50-200$ were allowed), the best-fitting model 
is marked with a square and the $1\sigma$ confidence region is
shaded. {\bf A high resolution version of this figure is available at
http://www.roe.ac.uk/$\sim$elt/N2paper\_images.html .}}
\end{figure}


\begin{figure}
\centerline{\psfig{figure=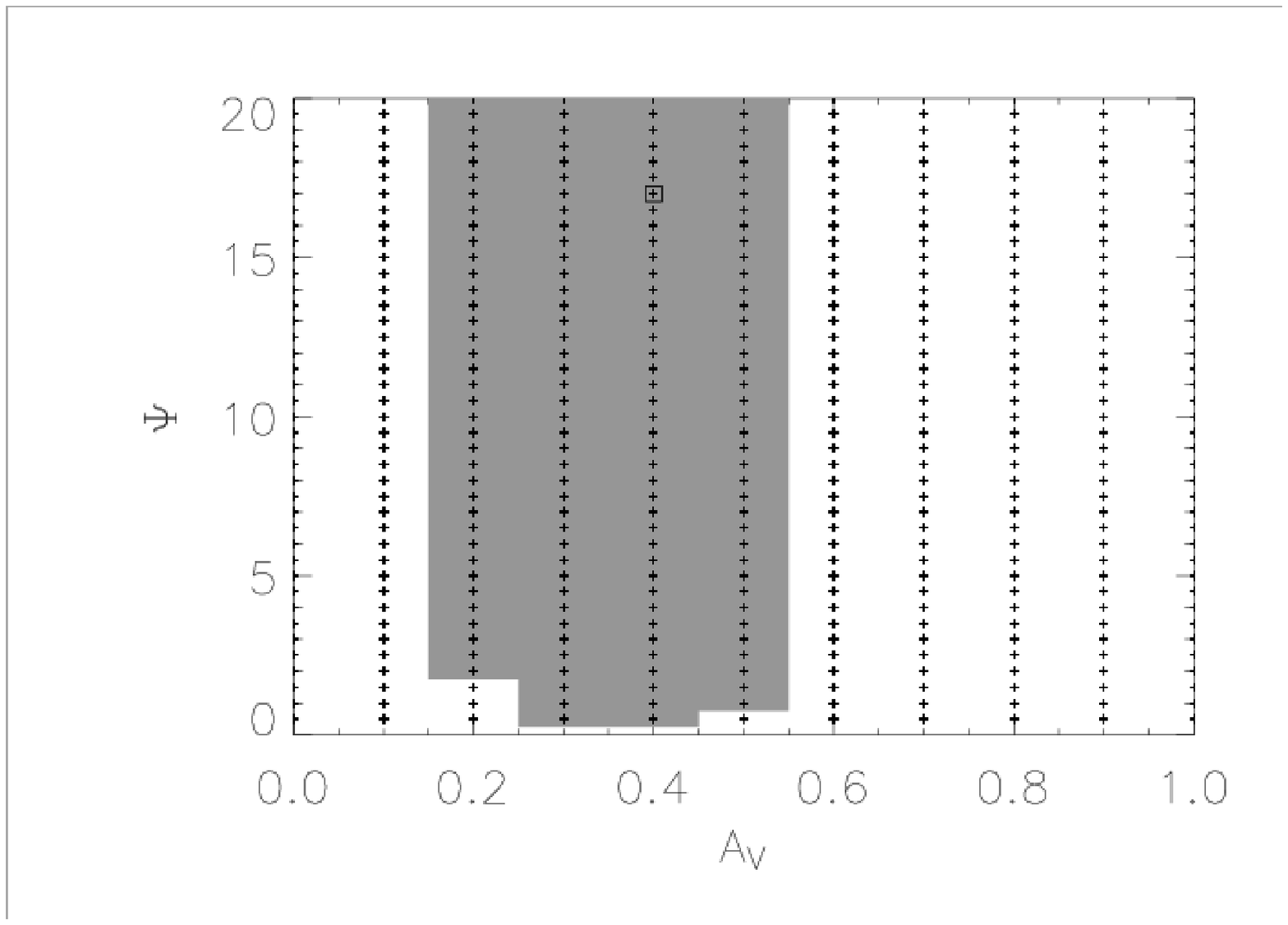,width=0.5\textwidth}}
\caption{\label{fig:av_chi}$1\sigma$ non--starburst model
parameter confidence region for
$A_V$ and $\psi$ for FN2 001. Possible models parameter combinations are shown
with small crosses (note $A_V=0-10$ and $\psi=0.5-20.0$ were allowed), the best-fitting model 
is marked with a square and the $1\sigma$ confidence region is
shaded. {\bf A high resolution version of this figure is available at
http://www.roe.ac.uk/$\sim$elt/N2paper\_images.html .}}
\end{figure}

\section{Radiative Transfer Models} \label{sec:models}
It was initially attempted to fit the data to the GRASIL model SEDs of
\citet{Silva98}. These provide six templates for nearby galaxies, three
starbursts (Arp 220, M82 and NGC 6090) and three sprials (M100, M51
and NGC 6946). However, these templates did not describe the data
well. Instead radiative transfer models of \citet{ERR03} and \citet{ERS2000} were employed 
to generate SED predictions for comparison with 
the data. These are able to model both cirrus and starburst emission
or a combination of the two. The models used are briefly described below.

Ultraviolet to submillimetre emission from stars
embedded in interstellar dust in galaxies known as the 'infrared
cirrus component' are modelled. The model consists of the input stellar radiation
field, an interstellar dust model and the radiative transfer treatment
of the interaction between the two which generates the output SED.

The input stellar radiation field is taken from the Galaxy Isochrone
Synthesis Spectral Evolution Library (GISSEL, \citealt{BC93}
). This gives the
radiation from a mass of stars from the ultraviolet to near-infrared
as a function of time.

Stars form in molecular clouds and at the early stages of their lives
remain inside them during which time their radiation is absorbed by
the dust and reprocessed to the infrared. During this phase of their
life the model uses the code of 
\citet{ERS2000} 
to compute the radiation emitted by stars and dust. The cloud
is assumed to disperse $7.2\times 10^7$ years after star formation. Before this, however, due to non-spherical
evolution of the cloud, a fraction of the starlight $f$ may have been
able to escape without dust absorption. This occurs a time $t_m$ after
star formation. 

The radiation field within the galaxy is modelled to be due to a large
number of randomly oriented molecular clouds whose average emission,
for stars in the age range $t_m$ to $7.2\times 10^7$ years, is
approximately
\begin{displaymath}
(1-f)S^{S}_{\nu}+fS^{\ast}_{\nu}
\end{displaymath} 
were $S_{\nu}^S$ is the emission from a giant spherical molecular
cloud \citep{ERS2000} 
and $S_{\nu}^{\ast}$ is the emission from the stellar population
\citep{BC93}. 
The emission of stars younger than $t_m$ is simply
$S_{\nu}^S$ and those older than $7.2\times 10^7$ years
$S_{\nu}^{\ast}$. The fraction of general starlight that escapes the galaxy unattenuated 
is also parameterised in a 'leak' variable. 

The dust temperature is determined by the intensity of the stellar
radiation field and is characterised in the model by the ratio of the
bolometric intensity of the radiation field to that in the stellar
radiation field in the solar neighbourhood \citep{MMP83}
, $\psi$. The star
formation rate, $\dot{\phi}_{\ast}(t)$, is assumed to have an exponential form with a time
scale, $\tau$, and a Salpeter IMF from $0.1-125$ M$_{\odot}$. 
\begin{displaymath}
\dot{\phi}_{\ast}(t)\propto e^{\frac{-t}{\tau}}
\end{displaymath}
In the case where a starburst is included the
star formation rate is modified to
\begin{displaymath}
\dot{\phi}_{\ast}^{SB}(t)\propto \dot{\phi}_{\ast}(t)+f_{SB}e^{\frac{-(t-t_{SB})}{\tau_{SB}}}
\end{displaymath}
where $f_{SB}$ is the ratio of the star formation rate at the peak of
the starburst to that at time 0, $t_{SB}$ is the age of the galaxy
at the start
of the starburst and $\tau_{SB}$ is the exponential timescale of the
starburst.
\setcounter{table}{1} 
\begin{table}
\caption{Parameter values}
\label{tab:params}
\begin{tabular}[h]{lcc}
 & starburst & non-starburst \\
\hline
$\tau$ & 8 Gyrs & 1 Myr \\
$t_{SB}$ & galaxy age$-t_m$ & - \\
galaxy age & \multicolumn{2}{c}{(universe age at z - 0.5 Gyrs)} \\
$\tau_{SB}$ & 40 Myrs & - \\
leak & 0.0 & 0.1 \\
f & 1.0 & 1.0 \\
$t_m$ & 20 Myrs & 3 Myrs \\
$A_V$ & 0--3 & 0--10 \\
$\tau_v$ & 50--200 & - \\
$\psi$ & 5 & 0.5-20.0 \\
$f_{SB}$ & 1--100 & - \\
\hline
\end{tabular}
\end{table}

The interstellar dust model of \citet{SK92} 
is used, which models the effects of both small grains and
PAHs. The opacity of the dust is characterised by $A_V$, which controls
the amount of UV to near-IR light absorbed by dust and re-emitted in
the IR and sub--mm and therefore the ratio of the luminosity in these
two bands. The model also has the ability to increase the dust
extinction of the starburst component of the galaxy separately via the
parameter $\tau_{v}$, values of $50,100,150,200$ are suppported. In
the starburst case a grid of models was produced over $f_{SB}$,
$A_V$ and $\tau_{v}$ and for non-starburst galaxies a grid of models
was produced over $A_V$ and $\psi$. The values of the parameters used
are given in Table \ref{tab:params}.

The effects of each of the model parameters on the resulting SED is
shown in Figure \ref{fig:mod_examples}. The solid black line shows an
SED produced using the fixed parameters ($\psi =5$, $\tau =8\times
10^{9}$, $t_m =20$ Myrs, $\tau_{SB} =40$ Myrs) and $f_{SB}=1$, $A_V=1$, and 
$\tau_v=100$ for comparison. A model with increased dust extinction of $A_V=3$ is shown by the black dotted 
line: this increases the emission at longer wavelengths and decreases
that in the optical, as would be expected. The dash-dot-dot-dot line shows 
a model with $t_m=0$ which dramatically changes the shape of the
far--infrared peak due to the reduction in the amount of dust
reprocessing of radiation from newly formed stars. However, any
significant amount of time spent within a molecular cloud produces a
far--infrared peak of shape similar to that of the other models
shown. The peak of the SED is also increased significantly with an
increase in the strength of the
starburst, a model with $f_{SB}=5$ is shown by the grey dashed
line. The other parameters do not affect the resulting SED vastly,
hence, there are degeneracies and a spread of parameter values
may result in a very similar SED. This is evidenced by models with
$\psi =10$ (dashed line), $\tau =8\times 10^{10}$ (dash-dot),
$\tau_{SB}=80$ Myrs (long dashes) and $\tau_v=200$ (solid grey
line) all having very similar shaped SEDs. The parameter values 
chosen were in
the mid-range of physically viable values as found by
\citet{ERS2000}. The $\tau_v$
range covers values known to fit normal starbursts $(\tau_v=50)$
\citep{ERS2000} to
heavily obscured starbursts such as Arp 220 $(\tau_v=200)$, this range 
was also used by \citet{Farrah} to fit a number of ULIRGs.

After fitting the models to the multiwavelength data, the parameter
values for models within the $1\sigma$ $\chi^2$ confidence limit of
the best-fitting model were examined.  The $1\sigma $ limit for starburst
model fitted SEDs typically encompasses $\tau_v \pm 50$, $A_V
\pm 0.5$ and $f_{SB}\pm 1$. There 
is also a correlation between $f_{SB}$ and $A_V$ such that models with
higher $f_{SB}$ values require low $A_V$ values to fit the data. This
is due to an increase in each of the parameters increasing the height
of the far--infrared peak (as shown in Figure \ref{fig:mod_examples}), 
therefore when both the parameters have
high values the height of the peak is over--estimated. We also note
that it is difficult to constrain $\tau_v$ since we do not have data
in the region of the spectrum which it most affects. For
non-starburst fitted SEDs the confidence region includes $A_V \pm 0.4$ 
and $\psi =0-20$. We also note that since the effect of $\psi$ on the
shape of the model SEDs is not marked, it is difficult to constrain
this parameter and the $1\sigma$ confidence region encompasses the
whole allowed parameter range. However, since it has little effect on
the shape of the model SEDs this 
will not affect the property estimates calculated from the models. Examples of these confidence regions 
for both the starburst (FN2 008) and non-starburst (FN2 001) galaxies
are  shown in Figures \ref{fig:fsb_av} to \ref{fig:av_chi}. 


\begin{figure*}
\centerline{\psfig{figure=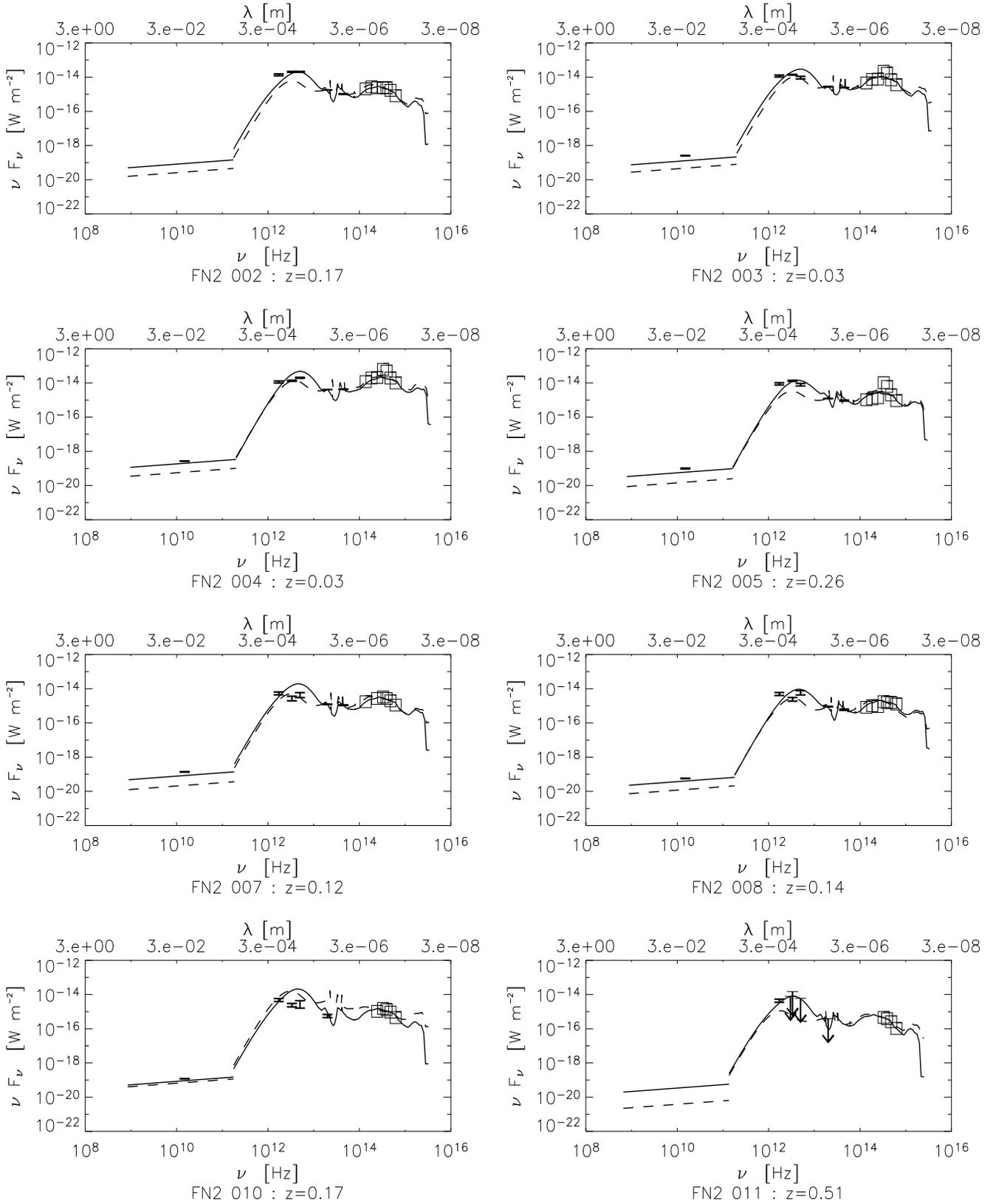,width=18cm,height=22cm}}
\caption{\label{fig:seds1}N2 ISO 175 ${\rm \mu m}$ sources fitted model
SEDs. Best-fitting starburst models (solid line) and cirrus models (dashed line)
are shown for each source. }
\end{figure*}


\begin{figure*}
\centerline{\psfig{figure=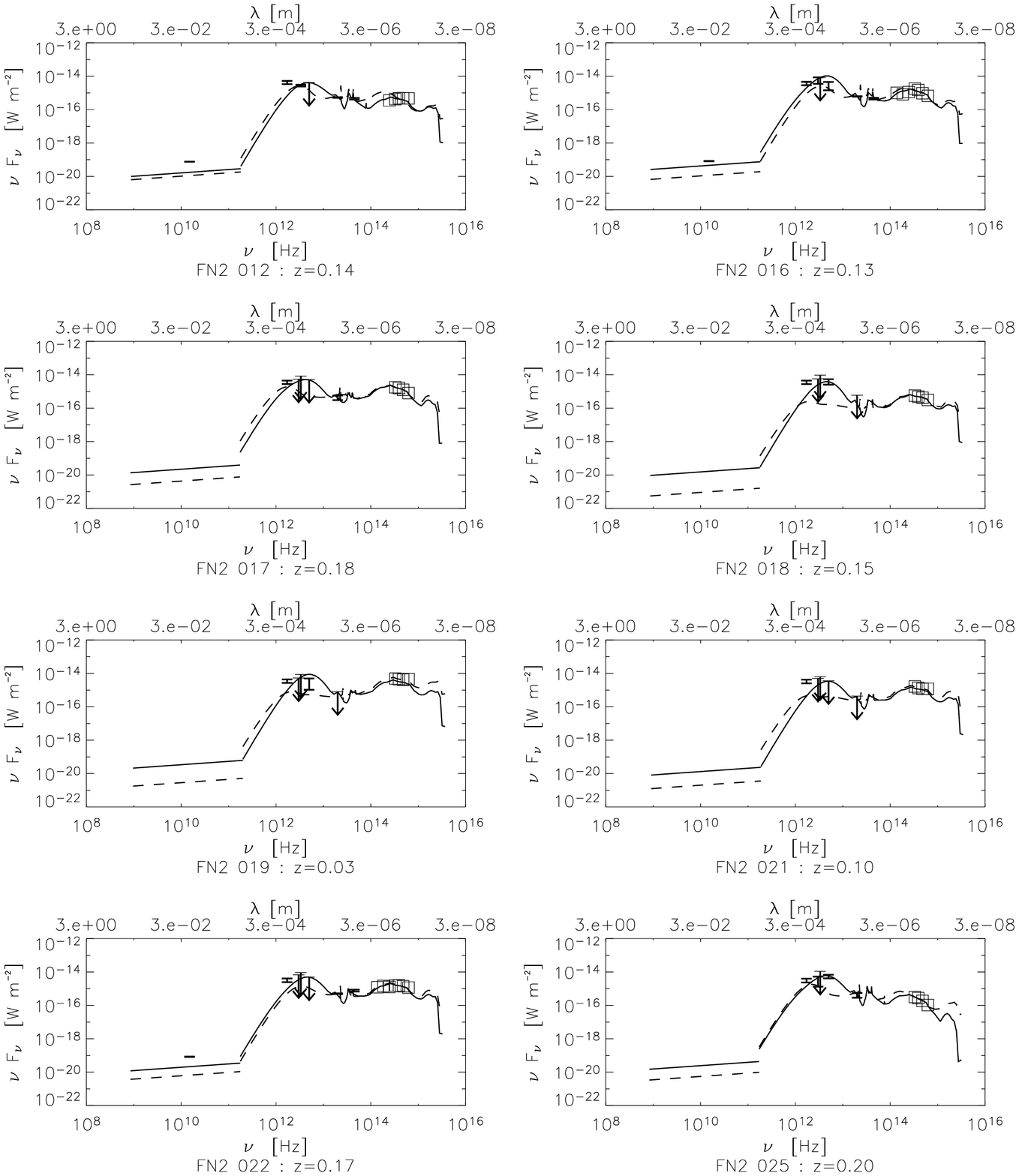,width=18cm,height=22cm}}
\contcaption{N2 ISO 175 ${\rm \mu m}$ sources fitted model
SEDs. Starburst models (solid line) and cirrus models (dashed line)
are shown. }
\end{figure*}


\begin{figure*}
\centerline{\psfig{figure=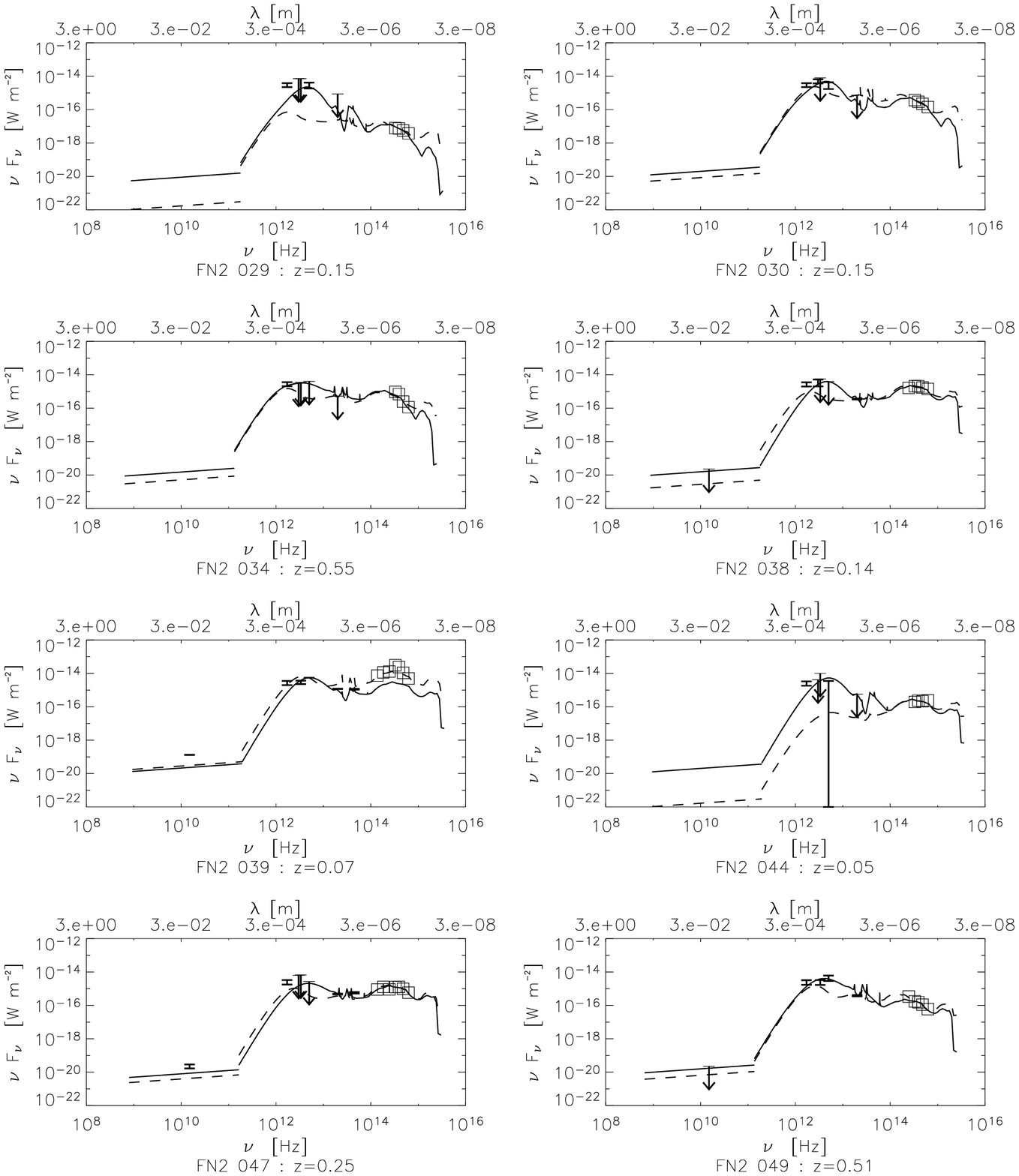,width=18cm,height=22cm}}
\contcaption{N2 ISO 175 ${\rm \mu m}$ sources fitted model
SEDs. Starburst models (solid line) and cirrus models (dashed line)
are shown. }
\end{figure*}

\section{Spectral Energy Distributions} \label{sec:SEDs}

We have compared the observed spectral energy distributions (SEDs)
with predictions from the radiative transfer models of Efstathiou and
Rowan-Robinson (2003) and \citet{ERS2000} described in Section \ref{sec:models}.
 
Spectroscopic redshifts were used in the fitting of the model
SEDs, where available from the ELAIS catalogue \citep{ELAIS} or elsewhere.  In other cases photometric redshifts were
drawn from the ELAIS catalogue, and calculated bespoke by ImpZ
\citep{Babbedge} for the remainder. We note that since photometric
redshifts are typically accurate to about 10 per cent in $(1+z)$, for the
spectroscopic redshifts we have available this implies an accuracy of
around 0.11 to 0.14. Therefore, given that the solutions are based on
a maximum of four optical bands, the
agreement between our photometric and spectroscopic redshifts is
reasonable, the largest margin of error being about 20 per cent in $(1+z)$.
It is also important to note that there are no catastrophic outliers
which can occur due to degeneracies in the solution space of
photometric codes. These small redshift errors do not affect the
fitted SEDs significantly and the resulting errors on derived
properties are much smaller than those incurred by the $1\sigma$ error 
on the fitted SED $\chi^2$ value which are given for all derived quantities.

\begin{figure*}
\centerline{\psfig{figure=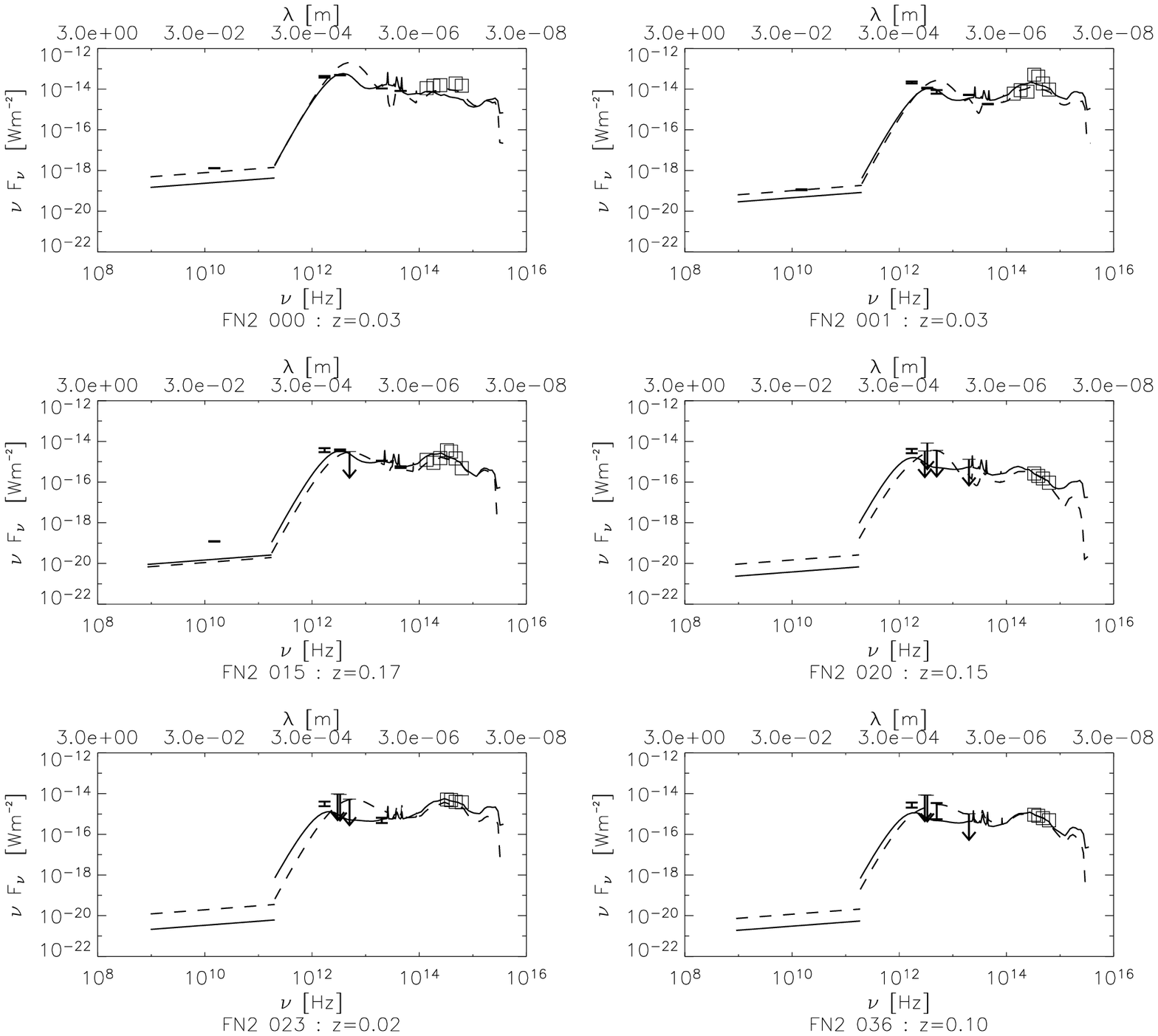,width=18cm,height=16.5cm}}
\caption{\label{fig:seds2}N2 ISO 175 ${\rm \mu m}$ sources fitted model
SEDs. Cirrus models (solid line) and starburst models (dashed line)
are shown. }
\end{figure*}

The N2 175 ${\rm \mu m}$ ISO source SEDs are shown in Figures \ref{fig:seds1} and
\ref{fig:seds2}. They appear to fall into two population categories,
the first being star bursting galaxies with their starburst region at a
high optical depth (fitted model parameters shown in Table
\ref{tab:sb_parameters} and SEDs plotted in Figure \ref{fig:seds1}) and the second
dust obscured galaxies with quiescent star formation (fitted model parameters
shown in Table \ref{tab:cirrus_parameters} and SEDs plotted in Figure \ref{fig:seds2}).

\begin{table}
\caption{Parameters for the starburst radiative transfer model SEDs fitted to
the N2 175 ${\rm \mu m}$ ISO source SEDs.}
\label{tab:sb_parameters}
\begin{tabular}{@{}lccc@{}}
\hline
FIRBACK & $A_V$ & $f_{SB}$ & $\tau_v$ \\
\hline
 FN2 002 & 0.4 & 2.0 & 150\\
 FN2 003 & 0.3 & 1.0 & 200\\
 FN2 004 & 0.0 & 1.0 & 200\\
 FN2 005 & 0.0 & 2.0 & 100\\
 FN2 007 & 0.2 & 2.0 & 200\\
 FN2 008 & 0.0 & 2.0 & 150\\
 FN2 010 & 0.3 & 3.0 & 200\\
 FN2 011 & 0.5 & 3.0 & 200\\
 FN2 012 & 0.0 & 3.0 & 100\\
 FN2 016 & 0.3 & 2.0 & 200\\
 FN2 017 & 0.4 & 1.0 & 200\\
 FN2 018 & 0.0 & 2.0 & 200\\
 FN2 019 & 0.0 & 1.0 & 200\\
 FN2 021 & 0.0 & 1.0 & 150\\
 FN2 022 & 0.1 & 1.0 & 150\\
 FN2 025 & 0.8 & 3.0 & 150\\
CFN2 029 & 1.1 & 10.0 & 200\\
CFN2 030 & 0.9 & 2.0 & 200\\
CFN2 034 & 1.0 & 1.0 & 50\\
CFN2 038 & 0.0 & 1.0 & 150\\
CFN2 039 & 0.0 & 1.0 & 100\\
CFN2 044 & 0.0 & 4.0 & 150\\
CFN2 047 & 0.0 & 1.0 & 50\\
CFN2 049 & 0.1 & 5.0 & 50\\
\hline
\end{tabular}
\end{table}

\begin{table}
\caption{Parameters for the quiescent radiative transfer model SEDs fitted to
the N2 175 ${\rm \mu m}$ ISO source SEDs.}
\label{tab:cirrus_parameters}
\begin{tabular}{@{}lcc@{}}
\hline
FIRBACK & $A_V$ & $\psi$ \\
\hline
 FN2 000 & 10.0 & 20.0 \\
 FN2 001 & 0.4  & 17.0 \\
 FN2 015 & 1.1  & 20.0 \\
 FN2 020 & 3.7  & 0.5 \\
 FN2 023 & 0.2  & 0.5 \\
CFN2 036 & 1.0  & 0.5 \\
\hline
\end{tabular}
\end{table}


\section{Star Formation Rates and Far--infrared Luminosities} \label{sec:sfrs}

\begin{figure}
\centerline{\psfig{figure=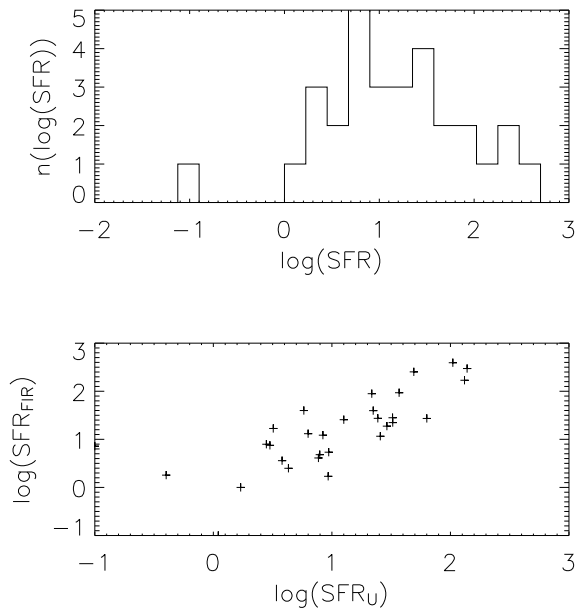,width=0.5\textwidth}}
\caption{\label{fig:sfr_hist}Top window:Far--infrared star formation rate distribution
($\textrm{M}_{\odot}\textrm{yr}^{-1}$) for the ISO 175 ${\rm \mu m}$
sources, see Table \ref{tab:sfrs_L}. 
Bottom window: Far--infrared starformation rate estimates vs. u-band
starformation rate estimates.}
\end{figure}

Star formation rates for the sources were calculated
using both an optical {\it U}-band estimate \citep{Cram} 
and a FIR
estimate \citep{Kennicutt}, 
given in equations \ref{eqn:sfr1} and
\ref{eqn:sfr2}.
\begin{equation} \label{eqn:sfr1}
{\rm SFR}_U({\rm M}\ge 5~{\rm M}_{\odot})=\frac{{\rm L}_U}{1.5\times 10^{22}~ \textrm{W Hz}^{-1}}100~\textrm{M}_{\odot}\textrm{yr}^{-1}
\end{equation}
\begin{equation} \label{eqn:sfr2}
{\rm SFR_{FIR}}=4.5\times 10^{-44} {\rm L_{FIR}}(\textrm{ergs s}^{-1})~\textrm{M}_{\odot} \textrm{yr}^{-1}
\end{equation}
Where L$_U$ is the {\it U}-band luminosity, estimated through interpolation
of the SED models and  ${\rm L_{FIR}}$ is the integrated infrared luminosity
over 8--1000 ${\rm \mu m}$. These both assume a Salpeter IMF. The
chosen IMF affects the estimated star formation rates since we
integrate over only a portion of the luminosity of the galaxy's
constituent stars. The proportion of stars emitting with luminosities
in the range in question will depend on the initial mass function,
since it describes the initial distribution of stellar luminosities over
the range, as discussed by \citet{Mann}. The star formation rates for
individual sources can be seen in Table \ref{tab:sfrs_L}. The distribution of
FIR star formation rates and a comparison between those estimated
using both the FIR and {\it U}--band luminosities can be seen in Figure \ref{fig:sfr_hist}. There
is a large spread in star formation rates with a number of galaxies
with SEDs suggestive of a modest starburst. This implies
that these galaxies are of low mass since their existing stellar
population is dominated by the small starburst at infrared
wavelengths. Using star
formation rates estimated for SED models falling in the $1\sigma$
$\chi^2$ range (see Section \ref{sec:models}) the {\it U}--band and FIR SFRs
are estimated to be accurate to 25 per cent and  50 per cent
respectively. It can be seen in the lower pannel of Figure
\ref{fig:sfr_hist} that the FIR star formation rate estimates are
significantly higher, typically by a factor of 4, than those using the U--band estimate. This is to
be expected for such highly obscured objects.

\begin{figure}
\centerline{\psfig{figure=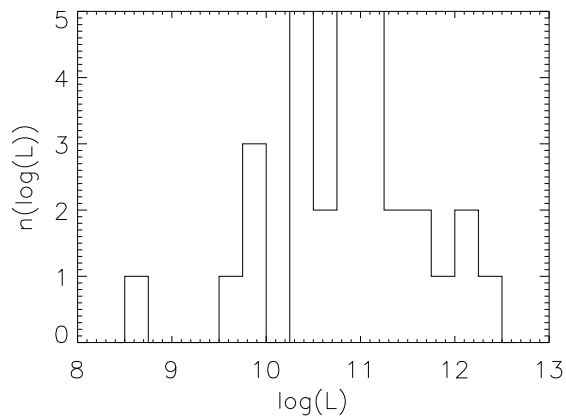,width=0.5\textwidth}}
\caption{\label{fig:l_hist}Distribution of far--infrared luminosities
(L$_{\odot}$) for the ISO 175 ${\rm \mu m}$ sources, see Table \ref{tab:sfrs_L}.}
\end{figure}

The integrated far--infrared luminosities over the
wavelength range 8-1000 ${\rm \mu m}$ were estimated. Results can be seen in
Table \ref{tab:sfrs_L} and a distribution in Figure
\ref{fig:l_hist}. These are estimated to be accurate to approximately
25 per cent.

\begin{table}
\caption{Star formation rates and FIR luminosities of the starburst
galaxies (first block) and cirrus galaxies (second block).}
\label{tab:sfrs_L}
\begin{tabular}{@{}lcccc@{}}
\hline
FIRBACK & {\it U}--band SFR & FIR SFR & $L_{FIR}$ & $z$\\
 & ${\rm M}_{\odot}{\rm yr}^{-1}$ & ${\rm M}_{\odot}{\rm yr}^{-1}$ & L$_{\odot}$ & \\
\hline
 FN2 002 &  36.9 &  93.1 & $4.87\times 10^{11}$ & 0.17\\
 FN2 003 &   3.8 &   3.6 & $1.83\times 10^{10}$ & 0.03\\
 FN2 004 &   9.4 &   5.4 & $2.85\times 10^{10}$ & 0.03\\
 FN2 005 & 131.4 & 169.1 & $9.18\times 10^{11}$ & 0.26\\
 FN2 007 &  22.3 &  39.5 & $2.07\times 10^{11}$ & 0.12\\
 FN2 008 &  24.4 &  27.3 & $1.46\times 10^{11}$ & 0.14\\
 FN2 010 &  21.7 &  88.9 & $4.68\times 10^{11}$ & 0.17\\
 FN2 011 & 104.8 & 391.9 & $2.04\times 10^{12}$ & 0.51\\
 FN2 012 &   6.3 &  13.1 & $7.07\times 10^{10}$ & 0.14\\
 FN2 016 &  12.6 &  25.6 & $1.33\times 10^{11}$ & 0.13\\
 FN2 017 &  32.5 &  28.1 & $1.41\times 10^{11}$ & 0.18\\
 FN2 018 &   8.4 &  12.2 & $6.53\times 10^{10}$ & 0.15\\
 FN2 019 &   1.7 &   1.0 & $5.22\times 10^{9}$ & 0.03\\
 FN2 021 &   7.9 &   4.8 & $2.59\times 10^{10}$ & 0.10\\
 FN2 022 &  32.6 &  22.3 & $1.18\times 10^{11}$ & 0.17\\
 FN2 025 &   5.8 &  39.7 & $2.07\times 10^{11}$ & 0.20\\
CFN2 029 &   0.1 &   7.2 & $3.79\times 10^{10}$ & 0.15\\
CFN2 030 &   3.2 &  16.8 & $8.60\times 10^{10}$ & 0.15\\
CFN2 034 & 138.2 & 296.6 & $1.48\times 10^{12}$ & 0.55\\
CFN2 038 &  25.6 &  11.6 & $6.19\times 10^{10}$ & 0.14\\
CFN2 039 &   7.7 &   4.1 & $2.23\times 10^{10}$ & 0.07\\
CFN2 044 &   0.4 &   1.8 & $9.41\times 10^{9}$ & 0.05\\
CFN2 047 &  63.2 &  27.2 & $1.49\times 10^{11}$ & 0.25\\
CFN2 049 &  49.1 & 252.8 & $1.39\times 10^{12}$ & 0.51\\
	&	&	&			&     \\
 FN2 000 &   2.8 &   7.9 & $3.97\times 10^{10}$ & 0.03\\
 FN2 001 &   9.3 &   1.7 & $8.37\times 10^{9}$ & 0.03\\
 FN2 015 &  29.1 &  18.8 & $9.58\times 10^{10}$ & 0.17\\
 FN2 020 &   3.0 &   7.5 & $2.56\times 10^{10}$ & 0.15\\
 FN2 023 &   1.1 &   0.1 & $3.87\times 10^{8}$ & 0.02\\
CFN2 036 &   4.3 &   2.5 & $8.82\times 10^{9}$ & 0.10\\
\hline
\end{tabular}
\end{table}

\section{Dust Temperatures and Masses} \label{sec:dust}

The dust in the galaxy radiative transfer models is at a
range of temperatures, however, we computed an emission-weighted estimate 
via the fitting of a grey body to the far--infrared portion
of the SED (from the peak of the SED to 1000 ${\rm \mu m}$) as follows:.
\begin{equation} \label{eqn:grey_body}
F_\nu \propto \frac{\nu^{\beta +3}}{e^{\frac{h\nu}{kT}}-1}
\end{equation}
In equation \ref{eqn:grey_body} T is the dust temperature which controls the position of
the maximum of the long wavelength SED, $\beta$ is the frequency dependence of
the dust grain emissivity and controls the gradient of the
long wavelength slope of the grey body curve. Fitting over this
wavelength range allows us to constrain both the temperature and
$\beta$, without reaching the far--infrared where the dust may go optically
thick and no longer be well described by a greybody model. The derived 
values
for the sources can be seen in Table \ref{tab:grey}. Using the method
described above (Section \ref{sec:sfrs}) for the starburst models these
are estimated to be accurate to T$\pm$3 K and $\beta \pm 0.5$, in the
non-starburst case these estimates are accurate to  T$\pm$14 K and $\beta \pm 0.1$.

\begin{figure}
\centerline{\psfig{figure=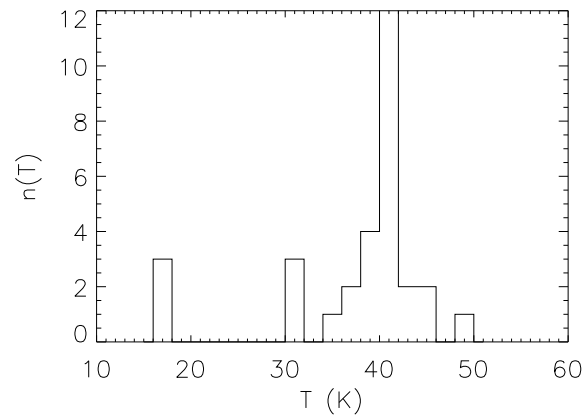,width=0.5\textwidth}}
\caption{\label{fig:t_hist}Distribution of dust temperatures for the
ISO 175 ${\rm \mu m}$ sources, see Table \ref{tab:grey}.}
\end{figure}

The distribution of the dust temperatures for the N2 175 ${\rm \mu m}$ ISO
source galaxies is shown in Figure \ref{fig:t_hist}. The starburst
galaxy dust temperatures are
slightly higher than those found by \citet{Sajina} ($\sim 20-30$ K) for 
a sample of N1
ISO 175 ${\rm \mu m}$ sources but have similar emissivity coefficients:
\citet{Sajina} found $\beta \sim 1.5-1.7$.

Dust masses were estimated using the submillimetre prescription
\citep{Hildebrand83} at a rest wavelength of 450 ${\rm \mu m}$:
\begin{eqnarray}
M_{dust}&=&\frac{S_{\nu_0}D_L^2}{(1+z)\kappa (\nu_r)B(\nu_r ,T_{dust})}\\
\kappa (\nu_r) &=& 0.067\left( \frac{\nu_r}{2.5\times
10^{11}}\right)^\beta
\end{eqnarray}
where $\nu_0$ and $\nu_r$ are the observed and rest frame frequencies
respectively, $S_{\nu_{0}}$ is the observed flux at $\nu_{0}$,
estimated through interpolation of the SED models, and $B(\nu_r,T_{dust})$ is
the Planck function in the rest frame. $T_{dust}$ and $\beta$ are the
dust temperature and dust grain emissivity estimated via the greybody
fitting. $\kappa (\nu_r)$ is known as the mass absorption coefficient
and has units of $\textrm{m}^2\textrm{kg}^{-1}$. This approach assumes 
that the galaxy is optically thin at submillimetre
wavelengths. Resulting dust masses, calculated at a wavelength of
450 ${\rm \mu m}$ (accurate to approximately 25 per cent), are shown in Table \ref{tab:grey}. We note that the dust
masses follow the trend of the far--infrared luminosities and star
formation rates, those with high SFRs and ${\rm L_{FIR}}$ having larger dust masses.

\begin{table}
\caption{Greybody parameters and dust masses for the starburst
galaxies (first block) and cirrus galaxies (second block).}
\label{tab:grey}
\begin{tabular}{@{}lccc@{}}
\hline
FIRBACK & T (K) & $\beta$ & ${\rm M_{dust}}({\rm M}_{\odot})$\\
\hline
 FN2 002 & 40 & 1.4 & $8.0\times 10^{7}$\\
 FN2 003 & 36 & 1.5 & $4.3\times 10^{6}$\\
 FN2 004 & 41 & 1.9 & $1.4\times 10^{6}$\\
 FN2 005 & 40 & 1.9 & $2.9\times 10^{7}$\\
 FN2 007 & 42 & 1.5 & $2.5\times 10^{7}$\\
 FN2 008 & 39 & 1.9 & $6.5\times 10^{6}$\\
 FN2 010 & 45 & 1.4 & $5.4\times 10^{7}$\\
 FN2 011 & 41 & 1.4 & $2.5\times 10^{8}$\\
 FN2 012 & 39 & 1.9 & $3.0\times 10^{6}$\\
 FN2 016 & 42 & 1.4 & $2.1\times 10^{7}$\\
 FN2 017 & 36 & 1.4 & $3.9\times 10^{7}$\\
 FN2 018 & 39 & 2.0 & $2.4\times 10^{6}$\\
 FN2 019 & 41 & 1.9 & $2.5\times 10^{5}$\\
 FN2 021 & 41 & 1.9 & $9.9\times 10^{5}$\\
 FN2 022 & 41 & 1.6 & $1.0\times 10^{7}$\\
 FN2 025 & 40 & 1.3 & $4.5\times 10^{7}$\\
CFN2 029 & 49 & 1.2 & $5.6\times 10^{6}$\\
CFN2 030 & 38 & 1.3 & $2.5\times 10^{7}$\\
CFN2 034 & 35 & 1.1 & $6.5\times 10^{8}$\\
CFN2 038 & 40 & 1.9 & $2.6\times 10^{6}$\\
CFN2 039 & 40 & 1.9 & $9.2\times 10^{5}$\\
CFN2 044 & 41 & 1.9 & $4.2\times 10^{5}$\\
CFN2 047 & 41 & 1.7 & $6.6\times 10^{6}$\\
CFN2 049 & 44 & 1.5 & $6.6\times 10^{7}$\\
	&	&	&		\\
 FN2 000 & 31 & 1.9 & $7.2\times 10^{6}$\\
 FN2 001 & 30 & 1.9 & $1.7\times 10^{6}$\\
 FN2 015 & 31 & 1.9 & $1.4\times 10^{7}$\\
 FN2 020 & 17 & 1.8 & $1.6\times 10^{8}$\\
 FN2 023 & 16 & 2.0 & $2.0\times 10^{6}$\\
CFN2 036 & 17 & 1.8 & $5.1\times 10^{7}$\\
\hline
\end{tabular}
\end{table}


\section{Evolutionary Models} \label{sec:evolution}
The detection of the far--infrared background in 1996 led to a
re--evaluation of galaxy evolution models. One of the earliest new models to
emerge was that of \citet{Burigana} which used the FIRB to place
constraints on models for the cosmic star formation history and
therefore the production of the background signal. This paper
concluded that stellar nucleosynthesis is able to account for a FIRB
intensity not far from the upper limit set by the COBE detection. It
also suggested substantially higher star formation rates at higher
redshifts than suggested by optical data.

It is now fairly well established that infrared--selected galaxies undergo
extremely high rates of evolution compared to measurements at other
wavelengths over the approximate redshift range $0<z<1$. Chary and
Elbaz (2001) models imply that about 80 per cent
of the cosmic infrared background was produced between $0<z<1.5$
compared to only 30 per cent of the 850 ${\rm \mu m}$ background. Also the dust
enshrouded star formation rate peaks at $z\sim 0.8$ at a value of $\sim
0.25~{\rm M}_{\odot}{\rm yr}^{-1}{\rm Mpc}^{-3}$ and at least 70 per cent of this star formation takes
place in infrared luminous galaxies with ${\rm L_{IR}}>10^{11}~{\rm L}_{\odot}$. In
fact $\sim 75$ per cent of 15 ${\rm \mu m}$ sources examined by \citet{E02} 
were found to be in this category of luminous infrared
galaxies and have star formation rates of $\sim 100~{\rm
M}_{\odot}{\rm yr}^{-1}$. The
comoving density of infrared light due to these galaxies was over 40
times larger at $z\sim 1$ than it is today.

Models of the cosmic infrared background produced by \citet*{Lagache}, 
that are able to fit number counts and unresolved
anisotropies imply that the luminosity function must change
dramatically with redshift. These also show rapid evolution of high
luminosity sources up to z=1. It has been suggested that this strong
evolution is due to bimodal star formation, one long lived quiescent,
passive phase and one enhanced starburst phase triggered by
interactions and mergers (e.g. \citealt{Lagache}, \citealt{F01}). \citet{F01} 
also suggest
that the interpretation of evolution with redshift could be biased due to an increase in the
probability of detecting a galaxy during an active phase caused by
both an increase of the rate of interactions, in part due to the
geometry of the expanding universe, and an increase of the infrared
luminosity due to more abundant fuel available in the past.

However, \citet{Chapman} 
point out that at $z<1$ the coldest and
most dusty galaxies will have the greatest flux densities for an
equivalent infrared luminosity. At $z>1$ the FIRBACK 175 ${\rm \mu m}$
detection will lie bluewards of the peak in the restframe black body
spectrum and dust temperatures of $<100$ K will no longer have
significant effects on the observed flux. Therefore at higher
redshifts, the flux limited
FIRBACK survey will be biased towards cooler galaxies.

Models such as \citet{F01} and \citet{CE01} do
not explicitly incorporate such cold luminous galaxies which are
preferentially selected in our survey for $z<1$. \citet{Chapman} examined
two of the 175 ${\rm \mu m}$ selected ISO sources and indentified them as
unusually cold ULIRGs with dust temperatures of $\sim 30$ K which is
significantly lower than the usual ULIRG dust temperature of $\sim
50$ K. Cold, dusty yet luminous objects suggest large masses of dust
heated at a moderate intensity. The existence of such sources may
seriously affect the interpretation of SCUBA sources if they
are numerous at high--z. Since many models rapidly evolve a hot ULIRG
population to the local luminosity function they may predict
a false redshift distribution if these cold sources are not considered.

\section{Discussion and Conclusions} \label{sec:conc}

We have found bright galaxy optical identifications for 31 out of the
55 N2 ISO 175 ${\rm \mu m}$
sources and stellar identifications for 2 sources. 17 of those with
galaxy optical IDs  were associated with sources from the band-mergerd ELAIS
catalogue. We then compard the spectral energy
distributions of the sources with
predictions from the radiative transfer models of \citet{ERR03} and \citet{ERS2000}.

22 of the 175 ${\rm \mu m}$ sources have not been confidently associated with 
optical counterparts and therefore their nature is still
unknown. However, only 4 of these sources are in the main FIRBACK
catalogue, the remaining 18 are listed in the complementary catalogue 
which contains detections at lower signal to noise and therefore they 
may just be spurious detections. However, these could also be due to sources
having multiple IDs or them being associated with heavily
extincted objects which were not in the optical catalogue.

24 galaxies have been classed as star bursting. They have a
range of far--infrared luminosities from $10^9~{\rm L}_{\odot}$ to
$10^{12}~{\rm L}_{\odot}$ corresponding to a range of star formation rates
from $\sim 0.1$ to $\sim 400~{\rm M}_{\odot}{\rm yr}^{-1}$. Although those at the
lower end of this range are not forming stars at a dramatic rate their
small starburst component is dominating over the existing
population at far--infrared wavelengths. This implies that the 175
${\rm \mu m}$ FIRBACK sample contains a
significant number of low mass galaxies. The majority of these
galaxies are spirals and a number are S0 galaxies.

Sources FN2 005, 011, CFN2 034 and CFN2 049 have ${\rm L_{FIR}}\sim
10^{12}~{\rm L}_{\odot}$ and high star
formation rates, we therefore suggest that these are in the low
redshift tail of the rapidly evolving ULIRG population. This is in
agreement with the findings of \citet{Sajina} who found one sixth of
their 175 ${\rm \mu m}$ sample to fall into this
category. We also suggest that those galaxies with moderate star formation rates and
far--infrared luminosities of $\sim 10^{11}~{\rm L}_{\odot}$ at
moderate z ($\ge 0.12$) are LIRGs (FN2 002, 007, 008, 010, 016, 017, 022, 025
and CFN2 047). Those with $z\le 0.12$ are more normal
galaxies perhaps with lower masses (FN2 003, 004, 012, 018, 019, 021,
CFN2 029, 030, 038, 039, 044).

The estimated dust temperatures of the starburst galaxies
are 30-49 K which is relatively low compared to that of standard ULIRGs
which have $T_D\sim50$ K. The most luminous of these galaxies also have 
large dust masses adding weight to the suggestion that cold,
dusty, luminous objects have large masses of dust heated at a moderate 
intensity \citep{Chapman}.

A further 6 sources have non-starburst SEDs. 
These sources have far--infrared luminosities
of $10^8 - 10^{10}~\textrm{L}_{\odot}$ and star formation rates of
$<30~\textrm{M}_{\odot}\textrm{yr}^{-1}$. We suggest that these are quiescently
star forming galaxies (FN2 000, 001, 015, 020, 023, 036). These
galaxies have low dust temperatures of $16-31$ K and dust masses of 
$10^6 - 10^8~\textrm{M}_{\odot}$. 3 of these galaxies are spirals, 1 is
an intereacting galaxy pair, 1 is either an
irregular or a spiral with tidal tail and the morphology
of the remaining source was unable to be deduced.

\citet{Lagache} predicted 62 per cent of 175 ${\rm \mu m}$ sources
would have $z<0.25$ and the remainder $0.8<z<1.2$. We find only one
175 ${\rm \mu m}$ source with
$S>180$ mJy with $z>0.25$ and therefore have $\gg62$ per cent with $z<0.25$.
However, our mix of higher redshift ULIRGs and lower redshift starburst galaxies is in
good qualitative agreement with the predictions of the \citet{Lagache}
starburst templates as adopted by \citet{Sajina}. \citet{Lagache} also state that to gain agreement
between their models and observations they require a cold local
population which is what has been found.

There has also been recent investigation into the submillimetre
properties of the local universe through the SCUBA Local Universe
Galaxy Survey \citep{DE01,DE00}. This survey found a population of
galaxies with ${\rm L_{FIR}}=10^{10}-10^{11}~{\rm L}_{\odot}$ and dust masses of
$10^7-10^8~{\rm M}_{\odot}$. The population was fitted with a two-component
dust model to discover cold dust temperatures of $\sim 21$ K and hot
dust temperatures of $30-50$ K. It is clear that the local 175 ${\rm \mu m}$ FIRBACK
population has very similar properties to these galaxies. The sources
of the far--infrared background are therefore offering a stepping stone 
between high redshift SCUBA galaxies and the local universe.

\section{Acknowledgements}
E. L. Taylor acknowledges support from a PPARC studentship.
Ph. H\'eraudeau acknowledges support from the EU TMR Network ``SISCO'' 
(HPRN-CT-2002-00316).
S. Mei thanks Marc Postman for insights on the galaxy morphological classification.
We thank an anonymous referee for a thorough reading of our
paper which significantly improved the presentation of its results.

\section{Appendix}
Notes on the multiwavelength associations for individual
sources. Images of each of the sources are shown in Figure \ref{fig:postages}.


\begin{figure*}
\caption{\label{fig:postages} Optical INT images of the N2 ISO
175 ${\rm \mu m}$
sources. FIRBACK source position marked with a diamond (axes show
distance from source position in arcsec), optical ID
marked with a cross, ELAIS positions marked with small circles, ELAIS
optical IDs marked with pluses. A 90 arcsec circle is plotted to show the
region in which optical IDs were searched for. {\bf Figure available
from http://www.roe.ac.uk/$\sim$elt/N2paper\_images.html.}}
\end{figure*}

\begin{description}
\item FN2 - 000: Identification with a bright optical galaxy
(${\rm P_{ran}}=0.026$, {\it r'} magnitude = 18.5, $z_{spec}=0.03$, irregular or
edge--on spiral with tidal tail). This is also
an ELAIS source (\verb|ELAISC15_J163734.4+405208|)
optical ID with detected emission at 15, 6.7, 90 ${\rm \mu m}$ and
radio detected emission at 20 cm.
There are two other ELAIS sources (\verb|ELAISC15_J163729.3+405248|, \verb|ELAISC15_J163731.3+405156|) within the FIRBACK error ellipse
with optical IDs with associated ${\rm P_{ran}}$ values of $0.604$ and $0.246$.  
\item FN2 - 001: Identification with a bright optical galaxy
(${\rm P_{ran}}=0.002$, {\it r'} magnitude = 18.3, $z_{phot}=0.10$, edge--on Sb/c). This is also
an ELAIS source (\verb|ELAISC15_J163507.7+405929|)
optical ID with detected emission at 15, 6.7, 90 ${\rm \mu m}$ and
radio detected emission at 20 cm. This also has an IRAS detection at $60 {\rm \mu m}$.
\item FN2 - 002: Identification with an optical galaxy
(${\rm P_{ran}}=0.078$, {\it r'} magnitude = 19.3). An ELAIS source ID (\verb|ELAISC15_J163608.1+410507|) has an
optical ID with the same galaxy, however, they are different entries
in the INT catalogue caused by the splitting up of bright objects by
the image analyser. The ELAIS ID has ${\rm P_{ran}}=0.086$,
$z_{spec}=0.17$, it is an irregulare galaxy with a tidal tail and
detected has emission at 15, 6.7 and 90 ${\rm \mu m}$. This also has an IRAS detection at 60 ${\rm \mu m}$.
There are two other ELAIS sources (\verb|ELAISC15_J163608.4+410529|, \verb|ELAISC15_J163611.6+410452|) within the FIRBACK error ellipse
with optical IDs with associated ${\rm P_{ran}}$ values of $0.885$ and $0.983$.
\item FN2 - 003: Identification with a bright optical galaxy
(${\rm P_{ran}}=0.001$, {\it r'} magnitude = 17.5, $z_{spec}=0.03$, Sa). This is also
an ELAIS source (\verb|ELAISC15_J163525.1+405542|)
optical ID with detected emission at 15, 6.7, 90 ${\rm \mu m}$ and
radio detected emission at 20 cm. This also has an IRAS detection at 60 ${\rm \mu m}$.
\item FN2 - 004: This ID is a small optical galaxy (${\rm P_{ran}}=0.079$, r
magnitude = 20.2). However, there is also present another large bright
galaxy (r magnitude = 16.4) which is mistakenly classified as a star
so is forcably not associated and assigned a ${\rm P_{ran}}$ of
1.0. Previous associations allowing ISO 175 ${\rm \mu m}$ sources to be associated
with stars chose this ID which matches with the optical ID of an ELAIS
source (\verb|ELAISC15_J163401.8+412052|) which has $z_{spec}=0.03$ and detected emission at 15, 6.7, 90 ${\rm \mu m}$ and
radio emission at 20 cm and has the morphology of a SBa. This also has an IRAS detection at 60 ${\rm \mu m}$.
\item FN2 - 005: Identification with bright optical galaxy (${\rm P_{ran}}=0.011$, r
magnitude = 18.7, $z_{spec}=0.26$, Sa). This is also an ELAIS source (\verb|ELAISC15_J163242.4+410847|)
optical ID with detected emission at 15, 6.7, 90 ${\rm \mu m}$ and
radio emission at 20 cm. This also has an IRAS detection at 60 ${\rm \mu m}$.  
There is another ELAIS source (\verb|ELAISC15_J163244.6+410911|) within the FIRBACK error ellipse with an
optical ID with an associated ${\rm P_{ran}}$ of $0.315$.
\item FN2 - 006: Identification with what appears on the image to be a
pair of interacting galaxies
(${\rm P_{ran}}=0.001$, {\it r'} magnitude = 18.2, no redshift). An ELAIS
source (\verb|ELAISC15_J163506.1+411038|) appears to be on the same object,
however, is does not have an
optical ID. There is another ELAIS source (\verb|ELAISC15_J163503.5+411137|) 
within the FIRBACK error ellipse. 
\item FN2 - 007: Identification with bright compact source with two nuclei (${\rm P_{ran}}=0.022$, r
magnitude = 18.0, $z_{spec}=0.12$). This is also an ELAIS source (\verb|ELAISC15_J163546.9+403903|)
optical ID with detected emission at 15, 6.7, 90 ${\rm \mu m}$ and
radio emission at 20 cm. This also has an IRAS detection at 60 ${\rm \mu m}$.
There is another ELAIS source (\verb|ELAISC15_J163545.5+403825|) within the FIRBACK error ellipse with an
optical ID with an associated ${\rm P_{ran}}$ of $0.998$.
\item FN2 - 008: Identification with an optical galaxy
(${\rm P_{ran}}=0.053$, {\it r'} magnitude = 18.4). However, the 175 ${\rm \mu m}$ flux
has been assigned in the ELAIS catalogue to a source (\verb|ELAISC15_J163548.0+412829|) with a different
optical ID. This ID has ${\rm P_{ran}}=0.084$ for association with the
FIRBACK source and is therefore a possible identification. The ELAIS
source has $z_{spec}=0.14$ and detected emission at 15, 6.7, 90 ${\rm \mu m}$ and
radio emission at 20 cm, it is an Sb galaxy. This also has an IRAS detection at 60 ${\rm \mu m}$.
\item FN2 - 009: This source does not have a confident optical
identification. It is best associated with a galaxy with
${\rm P_{ran}}=0.440$, this does not agree with the optical ID for the
ELAIS source (\verb|ELAISC15_J163359.2+405303|, ${\rm P_{ran}}=0.880$, {\it r'} magnitude=18.8) to which the 175 ${\rm \mu m}$ flux has been assigned, this
source also has detected emission at 15, 6.7, 90 ${\rm \mu m}$ and
radio emission at 20 cm. This source is confidently identified as a
star when stellar associations are allowed with ${\rm P_{ran}}=0.142$ (the
threshold for these associations is ${\rm P_{ran}}=0.4$), however, this
still does not agree with the ELAIS optical ID. 
\item FN2 - 010: Identification with
a pair of interacting galaxies, an irregular with a tidal tail and an Sa
(${\rm P_{ran}}=0.114$, {\it r'} magnitude = 18.8). An ELAIS source (\verb|ELAISR163536+411726|) ID has an
optical ID with the same galaxy, however, they are different entries
in the INT catalogue caused by the splitting up of bright objects by
the image analyser. The ELAIS source has $z_{spec}=0.17$ and detected emission at 90 ${\rm \mu m}$ and
radio emission at 20 cm. This also has an IRAS detection at 60 ${\rm \mu m}$.
\item FN2 - 011: Identification with an optical galaxy
(${\rm P_{ran}}=0.118$, {\it r'} magnitude = 19.8, $z_{phot}=0.51$, Sa?). However, the 175 ${\rm \mu m}$ flux
has been assigned in the ELAIS catalogue to a source (\verb|ELAISR163809+405839T|) with a different
galaxy ID (r magnitude = 18.6) and detected emission only in the far--infrared and radio. This has a ${\rm P_{ran}}$ value of 0.547 for association with the
FIRBACK source, it is therefore not an acceptable association.
\item FN2 - 012: Identification with an optical galaxy
(${\rm P_{ran}}=0.006$, {\it r'} magnitude = 18.5). However, the 175 ${\rm \mu m}$ flux
has been assigned in the ELAIS catalogue to a source (\verb|ELAISC15_J163412.0+405652|) with a different
galaxy ID (r magnitude = 20.0). This ID has ${\rm P_{ran}}=0.088$ for association with the
FIRBACK source and is therefore a possible identification. The ELAIS
source has $z_{spec}=0.14$ and detected emission at 15, 6.7, 90 ${\rm \mu m}$ and
radio emission at 20 cm, it is possibly an Sa.
There are two other ELAIS sources (\verb|ELAISC15_J163417.4+405710|, \verb|ELAISC15_J163417.9+405653|) within the FIRBACK error ellipse
with optical IDs with associated ${\rm P_{ran}}$ values of $0.990$ and $1.000$.
\item FN2 - 013: This source does not have a confident optical
identification. It is best associated with an optical galaxy
(${\rm P_{ran}}=0.260$, {\it r'} magnitude = 20.4). This is also an ELAIS source (\verb|ELAISC15_J163406.5+405106|)
optical ID with detected emission at 15 ${\rm \mu m}$.
\item FN2 - 014: This source appears in the ELAIS 90 and 175 ${\rm \mu
m}$ unassociated catalogue (\verb|ELAIS-FBK175_N2_014|). It does not have a confident optical identification.
\item FN2 - 015: Identification with a pair of interacting galaxies,
an Sa/b and an Sa, (${\rm P_{ran}}=0.015$, r
magnitude = 18.6, $z_{spec}=0.17$). This is also an ELAIS source (\verb|ELAISC15_J163607.7+405546|)
optical ID with detected emission at 15, 6.7, 90 ${\rm \mu m}$ and
radio emission at 20 cm.
\item FN2 - 016: Identification with an edge--on optical galaxy whose
morphological type is unclear
(${\rm P_{ran}}=0.119$, {\it r'} magnitude = 20.0). An ELAIS source (\verb|ELAISC15_J163423.9+405410|) ID has an
optical ID with the same galaxy, however, they are different entries
in the INT catalogue caused by the splitting up of bright objects by
the image analyser. This source has $z_{spec}=0.13$ and detected
emission at 15, 6.7 ${\rm \mu m}$ and radio emission at 20 cm. This also has IRAS
detections at 60 and 100 ${\rm \mu m}$.
There are two other ELAIS sources (\verb|ELAISC7_J163421+405413|, \verb|ELAISC7_J163423+405506|) within the FIRBACK error ellipse
one with an optical ID with ${\rm P_{ran}}=1.000$, the other has no optical ID.
\item FN2 - 017: Identification with a pair of interacting galaxies,
an Sa and an edge--on spiral,
(${\rm P_{ran}}=0.030$, {\it r'} magnitude = 18.5, $z_{phot}=0.18$). The 175 ${\rm \mu m}$ flux is assigned
to an ELAIS source (\verb|ELAISR163442+410759|) with a galaxy ID with associated ${\rm P_{ran}}=1.000$,
it is very faint and far from the source. There is, however, another
ELAIS source (\verb|ELAISR163445+410817|) present, with detected emission only in the radio, in the error ellipse without an optical ID which
is only about 5arcsec from our FIRBACK optical ID.
\item FN2 - 018: Identification with an optical galaxy
(${\rm P_{ran}}=0.037$, {\it r'} magnitude = 19.5, $z_{phot}=0.15$, SO/a). However, the 175 ${\rm \mu m}$ flux has been
assigned in the ELAIS catalogue to a source (\verb|ELAISC15_J163334.1+410139|) with a different galaxy
ID with ${\rm P_{ran}}=0.998$ (r magnitude = 20.5), it is therefore not a
possible ID. This also has an IRAS detection at 60 ${\rm \mu m}$.
There is another ELAIS source (\verb|ELAISR163333+410112|) within the FIRBACK error ellipse whose
galaxy ID has a ${\rm P_{ran}}=1.000$
\item FN2 - 019: Identification with an optical galaxy
(${\rm P_{ran}}=0.001$, {\it r'} magnitude = 18.1, $z_{phot}=0.10$, Sb). The ELAIS
source (\verb|ELAISR163716+404825|) to which the 175 ${\rm \mu m}$
flux has been assigned does not have an optical ID, it has detected emission at
$6.7 {\rm \mu m}$ and 20 cm. This also has an IRAS detection at 60 ${\rm \mu m}$.
\item FN2 - 020: Identification with an optical galaxy
(${\rm P_{ran}}=0.086$, {\it r'} magnitude = 20.2, $z_{phot}=0.15$). The ELAIS
source (\verb|ELAISC15_J163242.7+410627|) to which the
175 ${\rm \mu m}$ flux has been assigned is fainter (r magnitude = 22.1) and
further from the source than our ID and therefore has a higher
${\rm P_{ran}}$ value of 0.990.
\item FN2 - 021: Identification with an optical galaxy
(${\rm P_{ran}}=0.001$, {\it r'} magnitude = 17.6, $z_{phot}=0.10$). However, the flux of this FIRBACK source is below
the ELAIS catalogue $5\sigma $ limit of 223 mJy and therefore as it
has no associations at other ELAIS wavelengths it is omitted from the
catalogue. 
\item FN2 - 022: Identification with an optical galaxy
(${\rm P_{ran}}=0.019$, {\it r'} magnitude = 19.1). The ELAIS
source (\verb|ELAISC15_J163708.1+412856|) to which the 175 ${\rm \mu m}$ flux has been
assigned has an optical identification with a
bright optical galaxy with {\it r'} magnitude = 18.8, it has ${\rm P_{ran}}=0.140$
and is therefore a possible identification for the FIRBACK
source. This ELAIS source has $z_{spec}=0.17$, is an Sa and detected emission at 15, 6.7 ${\rm \mu m}$ and
radio emission at 20 cm.
There is another ELAIS source (\verb|ELAISC7_J163709+412832|) within the FIRBACK error elipse,
however, its optical ID is stellar and therefore it has ${\rm P_{ran}}=1.0$.
\item FN2 - 023: Identification with an optical galaxy
(${\rm P_{ran}}=0.018$, {\it r'} magnitude = 19.0, $z_{phot}=0.02$, Sb/c). However, the flux of this FIRBACK source is below
the ELAIS catalogue $5\sigma $ limit of 223 mJy and therefore as it
has no associations at other ELAIS wavelengths it is omitted from the
catalogue. 
\item FN2 - 024: This source does not have a confident optical
association. The flux of this FIRBACK source is also below
the ELAIS catalogue $5\sigma $ limit of 223 mJy and therefore as it
has no associations at other ELAIS wavelengths it is omitted from the
catalogue. 
\item FN2 - 025: Identification with an optical galaxy
(${\rm P_{ran}}=0.119$, {\it r'} magnitude = 20.2, $z_{phot}=0.20$, Sa).The ELAIS
source (\verb|ELAISC7_J163628+404757|) to which the 175 ${\rm \mu m}$ flux has been
assigned has 
optical ID classed as a galaxy, however, it looks stellar and 
has a ${\rm P_{ran}}$ of 0.974 so is not a possible ID.
There is another ELAIS source (\verb|ELAISC7_J163633+404749|) within the FIRBACK error elipse,
however, its optical ID is stellar and therefore it has
${\rm P_{ran}}=1.0$. This also has IRAS detections at 60 and
100 ${\rm \mu m}$.
\item FN2 - 026: This source does not have a confident optical ID. It
is associated with a galaxy with {\it r'} magnitude 22.4 with a
${\rm P_{ran}}=0.906$. There is an ELAIS source (\verb|ELAISC15_J163615.7+404759|) within the FIRBACK error
ellipse, however, it has a stellar optical ID and therefore has ${\rm P_{ran}}=1.0$.
\item CFN2 - 027: This source does not have a confident optical ID. It
is associated with a galaxy with {\it r'} magnitude 20.7 with a
${\rm P_{ran}}=0.305$. This does
not agree with the ELAIS source (\verb|ELAISR163703+412425|) optical ID which is a galaxy with ${\rm P_{ran}}=0.999$.
\item CFN2 - 028: This source does not have a confident optical
association. The flux of this FIRBACK source is also below
the ELAIS catalogue $5\sigma $ limit of 223 mJy and therefore as it
has no associations at other ELAIS wavelengths it is omitted from the
catalogue. 
\item CFN2 - 029: Identification with an optical galaxy
(${\rm P_{ran}}=0.070$, {\it r'} magnitude = 19.6, $z_{phot}=0.15$). The 175 ${\rm \mu m}$ flux has been
assigned to a fainter galaxy (\verb|ELAISR163418+410729|, {\it r'} magnitude = 22.2) which is further
from the FIRBACK position than our ID, it therefore has a high
${\rm P_{ran}}$ of 1.0. This also has an IRAS detection at 60 ${\rm \mu m}$.
There are three other ELAIS sources (\verb|ELAISR163419+410641|,
\verb|ELAISC15_J163421.4+410622|, \verb|ELAISR163422+410648|) within the FIRBACK error ellipse,
however, none of them have optical IDs.
\item CFN2 - 030: Identification with an optical galaxy
(${\rm P_{ran}}=0.086$, {\it r'} magnitude = 20.3, $z_{phot}=0.15$, S0/a). However, the flux of this FIRBACK source is below
the ELAIS catalogue $5\sigma $ limit of 223 mJy and therefore as it
has no associations at other ELAIS wavelengths it is omitted from the
catalogue. However, it has IRAS detections at 60 and 100 ${\rm \mu m}$. 
\item CFN2 - 031: This source does not have a confident optical
association. The flux of this FIRBACK source is also below
the ELAIS catalogue $5\sigma $ limit of 223 mJy and therefore as it
has no associations at other ELAIS wavelengths it is omitted from the
catalogue. 
\item CFN2 - 032: This source does not have a confident optical
association. The flux of this FIRBACK source is also below
the ELAIS catalogue $5\sigma $ limit of 223 mJy and therefore as it
has no associations at other ELAIS wavelengths it is omitted from the
catalogue. 
\item CFN2 - 033: This source does not have a confident optical ID. It
is associated with a galaxy with {\it r'} magnitude 21.8 with a
${\rm P_{ran}}=0.934$.
\item CFN2 - 034: Identification with an optical galaxy
(${\rm P_{ran}}=0.030$, {\it r'} magnitude = 19.4, $z_{phot}=0.55$, S0/a). However, the flux of this FIRBACK source is below
the ELAIS catalogue $5\sigma $ limit of 223 mJy and therefore as it
has no associations at other ELAIS wavelengths it is omitted from the
catalogue. 
\item CFN2 - 035: This source does not have a confident optical
association. The flux of this FIRBACK source is also below
the ELAIS catalogue $5\sigma $ limit of 223 mJy and therefore as it
has no associations at other ELAIS wavelengths it is omitted from the
catalogue. 
\item CFN2 - 036: Identification with an optical galaxy
(${\rm P_{ran}}=0.119$, {\it r'} magnitude = 19.4, $z_{phot}=0.10$, Sa). However, the flux of this FIRBACK source is below
the ELAIS catalogue $5\sigma $ limit of 223 mJy and therefore as it
has no associations at other ELAIS wavelengths it is omitted from the
catalogue. However, it has IRAS detections at 60 and 100 ${\rm \mu m}$. 
\item CFN2 - 037: This source does not have a confident optical ID. It
is associated with a galaxy with {\it r'} magnitude 22.1 with a
${\rm P_{ran}}=0.812$. When associations are allowed with all objects this
source is identified as a star with
${\rm P_{ran}}=0.183$. However, this still does not agree with the stellar optical
ID of the ELAIS source (\verb|ELAISR163812+405452|) assigned the 175 ${\rm \mu m}$ flux.
\item CFN2 - 038: Identification with an optical galaxy
(${\rm P_{ran}}=0.023$, {\it r'} magnitude = 19.3, $z_{spec}=0.14$, Sb/c). This is also the optical ID for
an ELAIS source (\verb|ELAISC15_J163431.5+412246|) with detected emission at 15 ${\rm \mu m}$ and has an IRAS detection at 100 ${\rm \mu m}$.
\item CFN2 - 039: Identification with an optical galaxy
(${\rm P_{ran}}=0.001$, {\it r'} magnitude = 17.2, $z_{spec}=0.07$, Sb). This is also the optical ID for
an ELAIS source (\verb|ELAISC15_J163613.6+404230|) with detected emission at 15, 6.7, 90 ${\rm \mu m}$ and
radio emission at 20 cm. This also has an IRAS detection at 60 ${\rm \mu m}$.
\item CFN2 - 040: This source does not have a confident optical ID. It
is best associated with a galaxy with {\it r'} magnitude 20.5 and
${\rm P_{ran}}=0.192$. The ELAIS source (\verb|ELAISC15_J163641.1+413131|) to which the 175 ${\rm \mu m}$ flux has
been assigned has a galaxy optical identification with
${\rm P_{ran}}=0.562$.
There is another ELAIS source (\verb|ELAISC15_J163648.1+413134|) within the FIRBACK error ellipse with a
galaxy optical ID with ${\rm P_{ran}}=0.393$.
\item CFN2 - 041: This source does not have a confident optical ID,
however, its best association agrees with the optical identification
of the ELAIS source (\verb|ELAISC15_J163433.6+405953|) to which the 175 ${\rm \mu m}$ flux has been
assigned. This galaxy has {\it r'} magnitude = 19.2 and ${\rm P_{ran}}=0.311$, the
ELAIS source has detected emission at 15 ${\rm \mu m}$ and 20 cm. 
There are two other ELAIS sources (\verb|ELAISR163427+405936T|, \verb|ELAISC15_J163430.1+410055|) within the FIRBACK error ellipse,
one has a galaxy optical ID with ${\rm P_{ran}}=0.998$ and the other does
not have an optical identification.
\item CFN2 - 042: This source does not have a confident galaxy optical
association. The flux of this FIRBACK source is also below
the ELAIS catalogue $5\sigma $ limit of 223 mJy and therefore as it
has no associations at other ELAIS wavelengths it is omitted from the
catalogue. 
\item CFN2 - 043: This source does not have a confident optical
identification when associations are only allowed with
galaxies (${\rm P_{ran}}=0.437$). However, when all objects are possible
associations it is matched with a star (${\rm P_{ran}}=0.056$) which is also
the ELAIS source (\verb|ELAISC7_J163546+404929|) ID to
which the 175 ${\rm \mu m}$ flux has been assigned. This ELAIS source has
detected emission at 6.7 ${\rm \mu m}$.
\item CFN2 - 044: Identification with an optical galaxy
(${\rm P_{ran}}=0.053$, {\it r'} magnitude = 20.0, $z_{phot}=0.05$, S0/a). The ELAIS
source (\verb|ELAISC15_J163730.4+404542|) to which the
175 ${\rm \mu m}$ flux has been assigned has a galaxy ID with a ${\rm P_{ran}}$
value of 0.995. This also has an IRAS detection at 60 ${\rm \mu m}$.
There is another ELAIS source (\verb|ELAISR163728+404533|) within the FIRBACK error ellipse,
however, this source does not have an optical ID.
\item CFN2 - 045: This source does not have a confident galaxy optical
association. The flux of this FIRBACK source is also below
the ELAIS catalogue $5\sigma $ limit of 223 mJy and therefore as it
has no associations at other ELAIS wavelengths it is omitted from the
catalogue. 
\item CFN2 - 046: This source does not have a confident optical
association. The flux of this FIRBACK source is also below
the ELAIS catalogue $5\sigma $ limit of 223 mJy and therefore as it
has no associations at other ELAIS wavelengths it is omitted from the
catalogue. 
\item CFN2 - 047: Identification with an optical galaxy
(${\rm P_{ran}}=0.070$, {\it r'} magnitude = 19.0). The 175 ${\rm \mu m}$ flux has been
assigned to an ELAIS source (\verb|ELAISC15_J163449.5+412048|) with a galaxy optical ID with
${\rm P_{ran}}=0.125$ which is therefore a possible association. This source 
has $z_{spec}=0.25$, is an S0/a and detections at 15, 6.7 ${\rm \mu m}$ and 20 cm.
There is another ELAIS source (\verb|ELAISC15_J163451.9+411944|) within the FIRBACK error ellipse with a
galaxy optical ID with ${\rm P_{ran}}=0.911$.
\item CFN2 - 048: This source does not have a confident optical
identification, is is best associated with a galaxy with
${\rm P_{ran}}=0.920$. The ELAIS source (\verb|ELAISC15_J163739.2+405643|) to which the 175 ${\rm \mu m}$ flux has
been assigned does not have an optical identification.
\item CFN2 - 049: Identification with an optical galaxy
(${\rm P_{ran}}=0.092$, {\it r'} magnitude = 20.3, $z_{phot}=0.18$). An ELAIS
source (\verb|ELAISC15_J163741.3+411913|) ID has an
optical ID with the same galaxy, however, they are different entries
in the INT catalogue caused by the splitting up of bright objects by
the image analyser. The ELAIS source has detected emission at 15 and
90 ${\rm \mu m}$ and is possible an Sa. This also has an IRAS detection at 60 ${\rm \mu m}$.
There is another ELAIS source (\verb|ELAISC15_J163738.9+411840|) within the FIRBACK error ellipse with a
galaxy optical ID with ${\rm P_{ran}}=0.911$.
\item CFN2 - 050: This source falls between two WFS chips.
\item CFN2 - 051: This source does not have a confident optical
identification, it is best associated with a galaxy with
${\rm P_{ran}}=0.866$. The ELAIS source (\verb|ELAISC15_J163616.2+411426|) to which the 175 ${\rm \mu m}$ flux has
been assigned has a stellar optical ID and therefore is not a possible
ID for the FIRBACK source.
\item CFN2 - 052: This source does not have a confident optical
identification, it is best associated with a galaxy with
${\rm P_{ran}}=0.466$. The ELAIS source (\verb|ELAISC15_J163403.0+410350|) to which the 175 ${\rm \mu m}$ flux has
been assigned has a galaxy optical ID with ${\rm P_{ran}}=0.962$.
There are two other ELAIS sources (\verb|ELAISC7_J163407+410219|, \verb|ELAISR163410+410326|) within the FIRBACK error ellipse one
with a stellar ID and one with a galaxy ID, however, both have ${\rm P_{ran}}=1.0$.
\item CFN2 - 053: This source does not have a confident optical
identification, it is best associated with a galaxy with
${\rm P_{ran}}=0.441$. The ELAIS source (\verb|ELAISR163621+412244|) to which the 175 ${\rm \mu m}$ flux has
been assigned has a galaxy optical ID with ${\rm P_{ran}}=0.989$.
\item CFN2 - 054: This source does not have a confident optical
identification. When associated with all optical objects it is
confidently associated with a star with ${\rm P_{ran}}=0.215$, the ELAIS
source (\verb|ELAISC7_J163658+411417|) assigned the 175 ${\rm \mu m}$ flux also has a stellar ID with
${\rm P_{ran}}=0.350$ which is below the limit of 0.4 and therefore a
possible association.
\end{description}

\end{document}